\documentclass[12pt,a4paper]{article}
\usepackage{amssymb}
\begin{document}

\title{\bf STUDIES OF CONCENTRATION AND TEMPERATURE DEPENDENCIES OF
PRECIPITATION KINETICS IN IRON-COPPER
 ALLOYS USING KINETIC MONTE CARLO AND
STOCHASTIC STATISTICAL SIMULATIONS}
\renewcommand{\abstractname}{}
\author{ K. Yu. Khromov, V. G. Vaks, and  I. A. Zhuravlev}
\date{}
\maketitle

\centerline{\it National Research Center "Kurchatov Institute",
123182 Moscow, Russia}

\centerline{\it Moscow Institute of Physics and Technology,
Institutskii per. 9, Moscow oblast, 141700 Russia}

\centerline{\it e-mail: vaks@mbslab.kiae.ru}

\begin{abstract}
\noindent{\bf Abstract} -- The earlier-developed $ab \ initio$ model
and the kinetic Monte Carlo method (KMCM) are used to simulate
precipitation in a number of  iron-copper alloys with different
copper concentrations $x$ and temperatures $T$. The same simulations
are also made using the improved version of the earlier-suggested
stochastic statistical method (SSM). The results obtained enable us
to make a number of general conclusions about the dependencies of
the decomposition kinetics in Fe-Cu alloys on $x$ and $T$. We also
show that the SSM describes the precipitation kinetics in a fair
agreement with the KMCM, and employing the SSM in conjunction with
the KMCM enables us to extend the KMC simulations to the longer
evolution times. The results of simulations seem to agree with
available experimental data for \hbox{Fe-Cu} alloys within
statistical errors of simulations and the scatter of experimental
results. Comparison of results of simulations to experiments for
some multicomponent \hbox{Fe-Cu}-based alloys enables us to make
certain conclusions about the influence of alloying elements in
these alloys on the precipitation kinetics at different stages of
evolution.
\end{abstract}


\section{INTRODUCTION}

Studies of the composition and temperature dependencies of the
precipitation kinetics in alloys, in particular, in multicomponent
Fe-Cu-based steels used in many industrial applications, attract
great attention \cite{Goodman-73}-\cite{Soisson-10}. For such
studies, it seems to be useful to have a reliable information about
the similar dependencies for more simple, binary alloys treated as
the reference systems. For example, discussing the precipitation
kinetics in two Fe-Cu-based steels, NUCu-140 and NUCu-170, Kolli et
al. \cite{KS-08,KWZS-08} compared their experimental results  to
those for several binaries  Fe-$x$Cu with $x$ between 1.34 and 1.5
at.\%, while the two steels considered had notably different content
of copper, 1.17 and 1.82\%, respectively. As many characteristics of
nucleation and growth, in particular, sizes and density of
precipitates, strongly vary with $x=x_{\rm Cu}$, such comparison can
be not quite adequate. The precipitation characteristics can also
significantly depend on temperature, while measurements of such
dependencies, particularly for the nucleation stage, often meet
difficulties \cite{KS-08,KWZS-08}.

To obtain the quantitative information about  the precipitation
kinetics, particularly in the course of nucleation and growth, one
can use simulations of these processes, if both the microscopic
model and the methods of simulations used can be considered as
reliable. For the Fe-Cu alloys, such reliable $ab \ initio$ model
has been developed by Soisson and Fu \cite{SF-07}, and their
detailed kinetic Monte Carlo simulations of nucleation and growth in
the Fe-1.34Cu alloy at $T$=773 K revealed a good agreement with the
available experimental data.

Therefore, the first aim of this work is to use the model and the
kinetic Monte Carlo method (KMCM) developed in  \cite{SF-07} to
study the concentration and temperature dependencies of the
precipitation kinetics in binary Fe-Cu alloys for a number of copper
concentrations $x$ and temperatures $T$, including those used by
Kolli et al. \cite{KS-08,KWZS-08} for the NUCu steels. Comparison of
results of these simulations to the available experimental data
enables us to assess reliability of the Soisson and Fu model at
different $x$ and $T$, and also to discuss the differences in the
precipitation kinetics between the Fe-$x$Cu binaries and the
multicomponent Fe-Cu-based alloys with the same $x=x_{\rm Cu}$ and
$T$.

The second aim of this work is to discuss possible applications of
the earlier-suggested stochastic statistical method (SSM)
\cite{KSSV-11} for simulations of precipitation kinetics in those
cases when the KMCM meets difficulties. Such difficulties arise, for
example, in treatments of the coarsening stage or at relatively low
temperatures $T\lesssim 300^{\rm o}$ C (typical for service of many
nuclear reactors) when the KMC simulations become time-consuming
\cite{SF-07,SF-08}. Unlike the KMCM, the SSM allows parallelizing
computer codes which can greatly accelerate computations, and this
method seems also to be suitable for various generalizations, for
example, for considerations of lattice misfit effects. However, in
treatments of nucleation, some oversimplified models have been used
in Ref. \cite{KSSV-11} which resulted in some fictitious breaks in
simulated temporal dependencies. In this work we describe the
improved version of the SSM free from these shortcomings and show
that this version describes the main characteristics of nucleation,
including the density and sizes of precipitates, in a good agreement
with the KMCM.  We also show that employing the SSM in conjunction
with the KMCM  enables us to extend the KMC simulations to the first
stages of coarsening.

In Sec. 2 we briefly discuss the model, the alloy states and the
methods used in our simulations. In Sec. 3 we describe the improved
version of the SSM employed in this work. The results of our
simulations are discussed and compared to the available experimental
data in Sec. 4. The main conclusions are summarized in Sec. 5.

\section{MODELS AND METHODS OF SIMULATIONS}

For our simulations we use  the $ab$ $initio$ model of Fe-Cu alloys
developed by Soisson and Fu  and described in detail in Refs.
\cite{SF-07,SF-08,KSSV-11}. Here we only note that this model uses
the following values of the binding energy between two copper atoms
and between a copper atom and a vacancy, $E^{bn}_{\rm CuCu}$ and
$E^{bn}_{{\rm Cu}v}$, for the $n$-th neighbors (in eV):
\begin{eqnarray}
&&E^{b1}_{\rm CuCu}=0.121-0.182T,\qquad E^{b2}_{\rm
CuCu}=0.021-0.091T,\nonumber\\
&&E^{b1}_{{\rm Cu}v}=0.126,\qquad E^{b2}_{{\rm Cu}v}=0.139.
\label{SF}
\end{eqnarray}
The high values of  $E^{bn}_{\rm CuCu}$ correspond to the strong
thermodynamic driving force for precipitation, while the strong
attraction of a vacancy to copper atoms results in the strong
vacancy trapping by copper precipitates discussed in detail in
\cite{SF-07}.

\begin{center}

\vbox{\noindent TABLE 1. The Fe-$x$Cu alloy states considered and
parameters of ``thermodynamic'' critical embryos for these states
calculated by method of Ref. \cite{DV-98}.
\vskip3mm
\begin{tabular}{|c|c|c|c|c|c|c|}
\hline
&&&&&&\\
Alloy state&$T$, {\rm K}&$x$, {\rm at}\%& $s$&$F_c/T$&
$N_c^{th}$&$R_c^{th}$, nm \\
\hline
&&&&&&\\
$A$&773&1.17&  0.247&5.58&15.4&0.353\\
&&&&&&\\
$B$&773&1.34&  0.285&4.38&14.3&0.344\\
&&&&&&\\
$C$&773&1.82&  0.393&2.30&12.2&0.326\\
&&&&&&\\
$D$&713&1.34&  0.352&2.47&10.7&0.312\\
&&&&&&\\
$E$&663&1.34&  0.425&1.36&8.2&0.286\\
&&&&&&\\
$F$&561&0.78&  0.387&1.11&5.3&0.247\\
&&&&&&\\
$G$&873&1.15&  0.163&12.6&28.3&0.433\\

\hline
\end{tabular}}
\end{center}
\vskip3mm

The  alloy states (below for short: states) used in our simulations
are listed in Table 1. The degree of supersaturation for each of
these states is characterized by the reduced supersaturation
parameter \,$s$\, introduced in Refs. \cite{KSSV-11,DV-98}:
\begin{equation}
s(x,T)=[x-x_b(T)]/[x_s(T)-x_b(T)] \label{s}
\end{equation}
where the lower index \,$s$\, or \,$b$\, corresponds to the spinodal
or the binodal.  Values  \,$s<1$\, correspond to the nucleation and
growth  evolution type, and \,$s>1$,\, to  spinodal decomposition.

The states  $A$ and $C$ in Table 1 have the same temperature $T$ and
the copper content $x$ as the above-mentioned steels NUCu-140 and
NUCu-170 studied by Kolli et al. \cite{KS-08,KWZS-08}.  For the
states $B$ and $G$, kinetics of precipitation under thermal aging
was investigated in Refs.
\cite{Goodman-73,Kampmann-86,Mathon-97,Shabadi-11}, while for the
state $F$, precipitation under neutron irradiation was studied by
Miller et al. \cite{Miller-03}.  For the states $D$ and $E$,
experimental data about precipitation kinetics are unknown to us,
and our simulations are made to study the temperature dependence of
this kinetics.

In Table 1 we also present some parameters of the ``thermodynamic''
critical embryos for the states considered:  the nucleation barrier
\,$F_c$, the total number of copper atoms within the embryo,
$N_c^{th}$,   and the critical radius \,$R_c^{th}$\, defined as the
radius of the sphere having the same volume as $N_c^{th}$ copper
atoms in the BCC lattice of $\alpha$-iron with the lattice constant
$a$=0.288 nm:
\begin{equation}
R_c^{th}=a(3N_c^{th}/8\pi)^{1/3}=0.142\,(N_c^{th})^{1/3}\,{\rm nm}.
\label{R_c}
\end{equation}
The characteristics of critical embryos shown in Table 1 and Fig. 1
are calculated by the statistical method of Dobretsov and Vaks
\cite{DV-98} with the use of the $ab$ $initio$ model by Soisson and
Fu \cite{SF-07} mentioned above and the pair cluster approximation
which is typically highly accurate, particularly for dilute alloys,
as both analytical studies \cite{VZhKh-10} and comparisons with
Monte Carlo simulations \cite{VKh-1}  show. Table 1 and Fig. 1
illustrate, in particular, the decreasing of the nucleation barrier
$F_c$ and the embryo size $N_c^{th}$ with increasing supersaturation
$s$. Table 1 also shows that for the given supersaturation $s$,
lowering temperature $T$ affects the critical embryo characteristics
stronger than increasing concentration $x$ \cite{KSSV-11}. As
discussed below, the $N_c^{th}$ and $R_c^{th}$ values presented in
Table 1 are usually close to those estimated in KMC simulations and
in experiments.
\begin{figure}
\caption{(color online \cite{arXiv}). Concentration profiles
\,$\Delta c(R)=c(R)-x$\, in thermodynamic critical embryos for the
alloy states considered, where $c(R)=c_{\rm Cu}(r)$ is the mean
copper concentration at the distance $R$ from the center of embryo.
\label{crit-embryos}}
\end{figure}

Methods of both KMC and SSM simulations used in this work are
described in detail in Refs. \cite{SF-07,KSSV-11}. Here we only
mention some details of computations. For both the KMC and the SSM
simulations we usually employed the simulation volume $V_s
=(64a)^3$;\, for the state $G$ with a low supersaturation we used
the bigger $V_s =(128a)^3$.\, In our KMC simulations we usually
followed $N^{KMC}_s\sim 10^{13}$ Monte Carlo steps, which  at $V_s
=(64a)^3$\, took about five months on a standard workstation. The
SSM simulations for $V_s =(64a)^3$\, took usually about a month on a
standard workstation (with no parallelization of codes made).

\section{IMPROVEMENTS OF STOCHASTIC STATISTICAL METHOD}

The  original version of the SSM is described in Ref.
\cite{KSSV-11}. To explain its improvements made in this work we
first present the necessary relations from \cite{KSSV-11}. Evolution
of a binary alloy is described by a stochastic kinetic equation
written in the finite difference form for a short time interval
\,$\delta t$:\,
\begin{equation} \delta c_i\equiv c_i(t+\delta t)-c_i(t)=\delta
c_{i}^d+\sum_{j_{nn}(i)}\delta n_{ij}^f\label{SKE}
\end{equation}
where \,$c_i$\, is the occupation of site \,$i$\, by a minority
(copper) atom averaged over some locally equilibrated vicinity of
this site, and the ``diffusional'' term \,$\delta c_{i}^d$\,
corresponds to the average atomic transfer for a certain effective
direct atomic exchange (DAE) model:
\begin{equation}
\delta c_{i}^d\{c_k\}=\sum_{j_{nn}(i)}M_{ij}2\sinh
[\beta(\lambda_j-\lambda_i)/2]\,\delta t.\label{delta n^d}
\end{equation}
Here $\lambda_i$ is the local chemical potential of site $i$ given
by Eq. (20) in \cite{KSSV-11}, and $M_{ij}$ is the generalized
mobility:
\begin{equation}
M_{ij}=\gamma_{\alpha h}^{eff}
\,b_{ij}^h\exp\,[\beta(\lambda_i+\lambda_j)/2]\label{M_ij-b}.
\end{equation}
The factor \,$b_{ij}^h$ in (\ref{M_ij-b}) is some function of local
concentrations $c_i$ given by Eq. (33) in  \cite{KSSV-11}, while
$\gamma_{\alpha h}^{eff}$ is an effective rate of exchanges between
neighboring minority and host atoms, $\alpha$ and $h$  (copper and
iron), which is proportional to the product of analogous rates
$\gamma_{\alpha{\rm v}}$ and $\gamma_{h{\rm v}}$ describing
exchanges between a vacancy and  an atom $\alpha$ and between a
vacancy and an atom $h$, respectively:
\begin{equation}
\gamma_{\alpha h}^{\rm eff}= \gamma_{\alpha{\rm v}}\gamma_{h{\rm
v}}\,\nu(t).\label{gamma_alpha h^eff}
\end{equation}
Here the function $\nu(t)$  defined by Eq. (36) in \cite{KSSV-11}
determines the ``rescaling of time'' between the original
vacancy-mediated exchange model and the effective DAE model used for
simulations. Temporal evolution of this DAE model is described by
the ``reduced time'' \,$t_r$\, having the meaning  of  a mean number
of atomic exchanges \,$\alpha\leftrightharpoons h$\, for the
physical time $t$. The reduced time $t_r$ is related to this time
\,$t$\, by the following differential or integral relations:
\begin{equation}
dt_r=\gamma_{\alpha h}^{\rm eff}dt,\quad \gamma_{\alpha h}^{\rm
eff}=\gamma_{\alpha{\rm v}}\gamma_{h{\rm v}}\nu (t),\quad t=
\int_0^{t_r}dt_r'/\gamma_{\alpha h}^{\rm eff}(t_r').\label{t-t_r}
\end{equation}
The form of the function $t(t_r)$ is discussed below. As mentioned
in Ref. \cite{VZh-12}, Eqs. \hbox{(\ref{SKE})-(\ref{t-t_r})} can be
derived not only for a simplified ``quasi-equilibrium'' model used
in \cite{KSSV-11} that disregards renormalizations of effective
interactions in a nonequilibrium alloy (whose importance for the
diffusion kinetics was noted by Nastar et al. \cite {Nastar-00}) but
also with allowing for these renormalizations.

The last term \,$\delta n_{ij}^f$\, in Eq. \,(\ref{SKE}) is the
fluctuative atomic transfer through the bond \,$ij$\, described by
the Langevin-noise-type method: each $\delta n_{ij}^f$ is treated as
a random quantity with the Gaussian probability distribution:
\begin{equation}
W(\delta n^f_{ij})=A_{ij}\exp [(-\delta
n^f_{ij})^2/2D_{ij}]\label{W}
\end{equation}
where \,$A_{ij}$\,  is the normalization constant. The dispersion
\,$D_{ij}$\, is related to the mobility \,$M_{ij}$\,  and the time
interval \,$\delta t$\, in Eq. (\ref{delta n^d}) by the
``fluctuation-dissipation'' type relation:
\begin{equation}
D_{ij}=\langle(\delta n_{ij}^f)^2\rangle=2M_{ij}\,\delta t.
\label{D_ij}
\end{equation}
As discussed in detail in \cite{KSSV-11}, for the non-equilibrium
statistical systems under consideration, the Langevin-noise-type
equations (\ref{SKE})--(\ref{D_ij}) should be supplemented by the
``filtration of noise'' procedure eliminating the short-wave
contributions to fluctuations \,$\delta n_{ij}^f$\, as these
contributions are already included in the diffusional term \,$\delta
c_{i}^d$\, obtained by the statistical averaging over these
short-wave fluctuations. Therefore, in the last term of Eq.
(\ref{SKE}), the full fluctuative transfer \,$\delta n_{ij}^f$\,
should be replaced by its long-wave part \,$\delta n_{ij}^{fc}$.\,
It can be made by introducing a proper cut-off factor \,$F_c({\bf
k})$\, in the Fourier-component \,$\delta n_{f\alpha}({\bf k})$\, of
the full fluctuation \,$\delta n_{ij}^f\equiv\delta
n_{\alpha}^f({\bf R}_{s\alpha})$\, where \,${\bf R}_{s\alpha}$\,
denotes the position of the \,$ij$\, bond center in the appropriate
crystal sublattice \,$\alpha$\, formed  by these centers
\cite{SPV-08}:
\begin{eqnarray}
&&\delta n_{ij}^f\to\delta n^{fc}({\bf R}_{s\alpha})=\sum_{\bf
k}\exp (-i{\bf
kR}_{s\alpha})\,\delta n^f({\bf k})\,F_c({\bf k})\nonumber\\
&&\delta n^f({\bf k})=\frac{1}{N}\sum_{{\bf R}_{s\alpha}}\exp (i{\bf
kR}_{s})\,\delta n^f({\bf R}_{s\alpha}) \label{delta N-fk}
\end{eqnarray}
where \,$N$\, is the total number of lattice sites (or atoms) in the
crystal. The cut-off factor \,$F_c({\bf k})$\,  for the BCC lattice
can be taken in the Gaussian-like form:
\begin{equation}
F_c^{\rm BCC }({\bf k})=\exp\, [-4g^2(1-\cos
\varphi_1\cos\varphi_2\cos\varphi_3)]\label{F_c-BCC}
\end{equation}
where \,$\varphi_{\nu}=k_{\nu}a/2$;\, $k_{\nu}$\, is the vector
\,${\bf k}$\, component along the main crystal axis $\nu$; and $a$\,
is the BCC lattice constant. At large \,$g^2\gg 1$,\, the expression
(\ref{F_c-BCC}) is reduced to a Gaussian \,$\exp\,(-k^2l^2/2)$\,
with \,$l=ga$.\,   Therefore, the reduced length \,$g=l/a$\,
characterizes the mean size of locally equilibrated subsystems.

This size, generally, varies with the aging time  \,$t$\, or $t_r$.
In particular, after creation of a supercritical precipitate, the
degree of local equilibrium in the adjacent region should increase
with respect to other regions where such precipitates are not born
yet. Therefore, after completion of nucleation at some reduced time
$t_{rN}$ (which can be estimated as the time for which the density
of precipitates reaches its maximum), the alloy should rapidly
approach the two-phase equilibrium, and the length \,$l=ga$\, should
become large, which should lead to a sharp decrease of the
fluctuative terms \,$\delta n^f=\delta n^{fc}$\, in Eqs. (\ref{delta
N-fk}) and (\ref{SKE}).

To describe this physical picture with the minimal number of model
parameters, the time dependence \,$g(t_r)$\, was approximated in
\cite{KSSV-11} by a simple one-parametric expression (71) with a
break at $t_r=t_{rN}$. In the present work we use for \,$g(t_r)$\, a
smooth two-parametric expression:
\begin{equation}
g(t_r)=g_0(1+t_r^2/t_0^2)\label{g-t_r}
\end{equation}
where $g_0$  and $t_0$ have the order of magnitude of the reduced
critical radius $R_c/a$ and the reduced nucleation time $t_{rN}$,
respectively.

\begin{figure}
\caption {(color online \cite{arXiv}). Evolution of the free energy
per copper atom, \,$F(t_r)$\,  ($a$), and  the density $d_p(t_r)$ of
supercritical precipitates or their total number $N_p(t_r)$ within
simulation  volume $V_s=(64a)^3$ ($b$), obtained  for the state $A$
in the SSM simulations with different \,$g_0$\, and $t_0$. Curves 1,
2, and 3 (red, green and blue online \cite{arXiv}) correspond to
$t_0=40$ and \,$g_0$\,=1.4, 1.35, and 1.45, respectively. Curves 4
and 5 (red on-line) correspond to $g_0=1.4$ and \,$t_0$\,=50 and 30,
respectively. Thin vertical line corresponds to $t_{rN}=18$.
\label{F_tg-A}}
\end{figure}

\vskip5mm
 \vbox{\noindent TABLE 2. Parameters  $g_0$ and
$t_0$ in (\ref{g-t_r}), reduced nucleation time $t_{rN}$,  and
maximum precipitate density $d_{max}$ (in $10^{24} \ {\rm m}^{-3}$)
for the alloy states considered.
 \vskip3mm
\begin{tabular}{|c|c|c|c|c|c|}
\hline
&&&&&\\
Alloy state&$g_0$&$t_0$&$t_{rN}$& $d_{max}^{SSM}$ & $d_{max}^{KMC}$  \\
\hline
&&&&&\\
$A$&1.4&40&$\sim 20$&3.4&4.0\\
&&&&&\\
$B$&1.4&30&$\sim 20$&5.4,\ 5.8&6.1\\
&&&&&\\
$C$&1.45&20&$\sim 10$&14.9&14.2\\
&&&&&\\
$D$&1.4&10&$\sim 10$&9.7&10.1\\
&&&&&\\
$E$&1.6&10&$\sim 10$&13.6&11.2\\
&&&&&\\
$F$&1.6&5&$\sim 5$&9.9&8\\
\hline
\end{tabular}}
\vskip5mm

To estimate values \,$g_0$\, and \,$t_0$\, within the SSM, we can
use the ``maximum thermodynamic gain'' principle discussed in detail
in \cite{KSSV-11}: we suppose that the ``most realistic'' values of
these parameters correspond to the minimum of the free energy of an
alloy after completion of nucleation. To illustrate this variational
method of estimating \,$g_0$\, and $t_0$, in Fig. 2  we present the
SSM-simulated temporal dependencies of the free energy per copper
atom for the state $A$ in Table 1 at several \,$g_0$\, and \,$t_0$\,
close to their ``optimal'' values. The free energy \,$F=F(c_i)$\,
was calculated according to Eq. (24) in \cite{KSSV-11} with the
simulated $c_i=c_i(t_r)$ dependencies. For simplicity, the initial
state was taken uniform: \,$c_i(0)=c$\,=const, thus the initial
increase of \,$F$\, at \,$t_r\lesssim 0.3\,\,t_{rN}$\, seen in Fig.
2$a$ is related just to switching-on fluctuations at \,$t_r=0$.\,
Fig. 2$a$ shows that the influence of this spurious increase of $F$
ceases only at $t_r\gtrsim 2t_{rN}$, while at too long $t_r\gtrsim
4\,t_{rN}$,\, the fluctuations are effectively switched-off
according to Eqs. (\ref{delta N-fk})-(\ref{g-t_r}). Therefore, to
estimate parameters \,$g_0$\, and \,$t_0$,\,  we usually consider
the interval $2t_{rN} \gtrsim t_r\gtrsim 4\,t_{rN}$. For the most of
states considered, the free energy \,$F(t_r)$\, has here a distinct
minimum with respect to \,$g_0$\, and $t_0$, as illustrated by Fig.
2$a$. Fig. 2$b$ illustrates sensitivity  of the precipitate density
$d_p$ to the choice of $g_0$ and $t_0$.  The resulting scatter in
simulated $d_p(t_r)$  usually lies within statistical errors of
simulations.

In Table 2 we present the values of \,$g_0$\, and $t_0$ in
(\ref{g-t_r}) estimated as described above. For the state $G$ the
simulations are time-consuming, thus such estimates have not been
made in this work. To assess adequacy of our SSM simulations, in the
two last columns of Table 2 we compare the values of the maximum
density $d_{max}$ of supercritical precipitates obtained in the SSM
simulations to those obtained in the KMC simulations.  For the state
$B$ in Table 2, two values \,$d_{max}^{SSM}$\, correspond to the two
simulations (shown in Fig. 6)  with the different random number
sets. We see that the SSM and the KMC results for $d_{max}$ agree
within statistical errors of simulations.

The precipitation kinetics is usually characterized by the density
and the mean radius of supercritical precipitates, $d_p(t)$ and
$R(t)$, defined by the following relations:
\begin{eqnarray}
&&d_p(t)=\sum_k\nu_k(t)/V_s, \nonumber\\
&&R(t)=\sum_k\nu_k(t)R_k\Big/\sum_k\nu_k(t). \label{d_p-R_p-def}
\end{eqnarray}
Here $\nu_k(t)$ is the number of clusters containing  $k$ copper
atoms, $V_s$ is the simulation volume, $R_k$  is defined similarly
to Eq. (\ref{R_c}): $R_k=a(3k/8\pi)^{1/3}$, and sums over $k$ in
(\ref{d_p-R_p-def}) include only clusters with $k\geq N_c$ where
$N_c$ is the ``critical'' size chosen. As discussed in
\cite{SF-07,KSSV-11} and below, the exact choice of this size (if
reasonable) usually does not significantly affect the $d_p(t)$ and
$R(t)$ values in (\ref{d_p-R_p-def}). Therefore, we take $N_c$ close
to its ``thermodynamic'' value $N_c^{th}$ in Table 1 putting
$N_c$=15, 15, 12, 11, 8, 5 and 28 copper atoms for the state $A$,
$B$, $C$, $D$, $E$, $F$ and $G$, respectively. Temporal dependencies
$d_p(t_r)$ and $R(t_r)$ obtained in our SSM simulations with such
$N_c$ are illustrated by Fig. 3 for the state $A$; for the rest
states considered, these dependencies are similar. Fig. 3
illustrates the sequence of four well-defined stages of
precipitation accepted in the classical theory of nucleation (see,
e. g., \cite{SM-00}): incubation, nucleation, growth and coarsening,
as well as the presence of an ``intermediate'' stage between growth
and coarsening (seen also in the simulations of Soisson and Martin
\cite{SM-00} and discussed in detail by Barashev et al.
\cite{Barashev-04}) which corresponds to the beginning of Ostwald
ripening when the bigger precipitates already start to grow due to
dissolving the smaller ones but the latter do not disappear yet. For
brevity, this intermediate stage will be called ''pre-coarsening''.
\begin{figure}
\caption {(color online \cite{arXiv}). Evolution of the density of
precipitates, $d_p(t_r)$   ($a$), and their mean radius $R(t_r)$
($b$) obtained in the SSM simulations for the state $A$ in Table 1.
Dotted vertical lines correspond to the values $t_{ir}$ in Eqs.
(\ref{t_ri}).\label{d_p-R_p-A}}
\end{figure}

Let us now discuss the  ``rescaling of time'' function $t(t_r)$  in
Eq. (\ref{t-t_r}) determined by the temporal dependence of the
effective direct exchange rate $\gamma_{\alpha h}^{eff}=\gamma_{\rm
CuFe}^{eff}(t_r)$.\,  For simplicity, this dependence was
approximated in \cite{KSSV-11} by a simple two-parametric expression
(77) which included sharp breaks at some $t_r$. More adequate
interpolations for $t(t_r)$ can be obtained from comparison of the
SSM and the KMC results for the density and the mean size of
precipitates, that is, from an approximate solution of two
equations:
\begin{equation}
d_p^{KMC}(t)=d_p^{SSM}(t_r), \qquad R_{KMC}(t)=
R_{SSM}(t_r).\label{d_p-R_p}
\end{equation}
Our estimates of functions $t(t_r)$  for which  both equations
(\ref{d_p-R_p}) are satisfied with a reasonable accuracy showed that
for all alloy states considered, these functions have a similar form
characterized by the presence of four intervals of $t_r$ with an
approximately constant value of the derivative $D=d\ln t/d\ln
t_r\propto 1/\gamma_{\rm CuFe}^{eff}$ within each interval. These
intervals $(i)$ are determined by the inequalities:
\begin{eqnarray}
&&(1)\  t_r<t_{r1},\qquad (2)\ t_{r1}<t_r<t_{r2},\nonumber\\
&&(3)\ t_{r2}< t_r<t_{r3},\qquad (4)\ t_{r3}<t_r, \label{t_ri}
\end{eqnarray}
where the point $t_{r1}$,  $t_{r2}$ or  $t_{r3}$ approximately
separates the stages of nucleation and growth, growth and
pre-coarsening, or pre-coarsening and coarsening, respectively, as
illustrated by Fig. 3. Therefore, within each interval $(i)$ we use
a simple power interpolation
\begin{equation}
t=A_it_r^{D_i}\label{t-t_r_D_i}
\end{equation}
with the values of parameters $t_{ri}$, $D_i$ and $A_i$  given in
Table 3. For the state $F$ with a relatively low $T=561$ K, our KMC
simulations (illustrated by Fig. 11 below) are time consuming and
reach only the growth stage, thus the interpolation
(\ref{t-t_r_D_i}) has not been constructed  for this state.

Functions $t(t_r)$ obtained are shown in Fig. 4. To compare the
precipitation parameters at different temperatures $T$ for which the
equilibrium vacancy concentration $c_v^{eq}(T)$ can be very
different, in Fig. 4 we use the ``scaled'' time $t_s(t_r)$ defined
as
\begin{equation}
t_s(t_r)=t(t_r)\,c_v^{eq}(T)/c_v^{eq}(773\,{\rm K}),\label{t_s-t_r}
\end{equation}
to cancel  the standard scaling factor $1/c_v^{eq}$ in the $t_s$
value. For the  $c_v^{eq}(T)$ we use the Soisson and Fu estimate
\cite{SF-07}:
\begin{equation}
c_v^{eq}(T)=\exp\,(-\varepsilon_v^{for}/T),\qquad\varepsilon_v^{for}=2.18
\,{\rm eV}.\label{c_v^eq}
\end{equation}
The SSM values $d_p^{SSM}(t)$ and $R_{SSM}(t)$ found using Eqs.
(\ref{t-t_r_D_i}) and Table 3 are compared to the appropriate KMC
results in Figs. 5-10. Note that the interpolation (\ref{t-t_r_D_i})
includes some minor breaks at points $t_r=t_{ri}$ which are seen in
Fig. 4. These breaks can be removed by introducing some smooth
matching of two lines (\ref{t-t_r_D_i}) adjacent to each point
$t_{ri}$, which also leads to the better agreement between the SSM
and the KMC results in Figs. 5-10. However, to avoid introducing too
many interpolation parameters, in this work we use a more simple
interpolation (\ref{t-t_r_D_i}) which seems to provide a
sufficiently adequate description of precipitation.

As  discussed in Ref. \cite{KSSV-11}, the significant changes of
derivatives $D_i\sim 1/\big(\gamma_{\rm CuFe}^{eff}\big)_i$\,
between different intervals $i$ seen in Fig. 4 can be related to the
strong vacancy trapping by copper precipitates and to the changes of
scale of this trapping in the course of precipitation. In
particular, the relatively low values of derivatives $D_3$\, for the
pre-coarsening stage can be related to the strong vacancy trapping
for this stage (illustrated by Fig. 6a in \cite{SF-07}) which should
result in a sharp increase of the effective rate  $\gamma_{\rm
CuFe}^{eff}$ in Eq. (\ref{t-t_r}) \cite{KSSV-11}. Note also
similarity of the curves $t_s(t_r)$ for different alloy states in
Fig. 4, as well as rather smooth variations of parameters $D_i$ and
$A_i$ in Table 3 under changes of concentration $x$ and temperature
$T$ corresponding to these different states. It seems to reflect a
great similarity of the vacancy trapping effects within each
interval $i$ considered for the different alloy states. This
similarity can be used for various SSM-based extrapolations of KMC
simulations, in particular, for the SSM-based simulations of
precipitation in Fe-Cu alloys at most different concentrations $x$
and temperatures $T$ with the use for the parameters $t_{ri}$, $D_i$
and $A_i$ in Eqs. (\ref{t_ri}) and (\ref{t-t_r_D_i}) some
interpolations between their values presented in Table 3. This
similarity is also used for the extrapolations of KMC simulations to
the first stages of coarsening discussed below.

\begin{figure}
\caption {(color online \cite{arXiv}). The scaled physical time
$t_s(t_r)$ defined by Eqs. (\ref{t-t_r_D_i}) and (\ref{t_s-t_r})
versus the reduced time $t_r$. Each curve corresponds to the alloy
state indicated by a symbol near this curve.\label{t_t_r-curves}}
\end{figure}

\begin{center}

\vbox{\noindent TABLE 3. Values of parameters  $t_{ri}$,  $D_i$ and
$A_i$ (in hours) in Eqs.   (\ref{t_ri}) and (\ref{t-t_r_D_i}).
 \vskip3mm
\begin{tabular}{|c|ccc|cccc|cccc|}
\hline
&&&&&&&&&&&\\
Alloy state&$t_{r1}$&$t_{r2}$&$t_{r3}$&$D_1$&$D_2$&$D_3$&$D_4$&$A_1$&$A_2$&$A_3$&$A_4$\\
\hline
&&&&&&&&&&&\\
$A$&18&506&5500&0.7&0.23&0.04&0.3&0.061&0.24&0.77&0.08\\
&&&&&&&&&&&\\
$B$&7.4&192&7240&1.1&0.33&0.11&0.4&0.021&0.097&0.30&0.023\\
&&&&&&&&&&&\\
$C$&2.9&70&3000&1.85&0.23&0.12&0.35&0.013&0.071&0.11&0.018\\
&&&&&&&&&&&\\
$D$&5&60&2300&1&0.23&0.08&0.27&0.027&0.14&0.26&0.026\\
&&&&&&&&&&&\\
$E$&3.2&60&2200&1.5&0.19&0.08&0.3&0.015&0.16&0.25&0.012\\
\hline
\end{tabular}}
\end{center}
\vskip3mm

To conclude this section we note that the simplified treatment of
fluctuations based on Eqs. (\ref{W})-(\ref{g-t_r}) can adequately
describe nucleation and growth only when these two processes are
sufficiently separated from each other, so that switching off
fluctuations after completion of nucleation implied by Eq.
(\ref{g-t_r}) can be appropriate. Figs. 5-10 show that this
condition is more or less satisfied for the first five states in
Table 1. At the same time, for the states $F$ and $G$, that is, at
low temperatures $T\lesssim 300^{\rm o}$ C or at low
supersaturations $s\lesssim 0.2$, the nucleation, growth and
pre-coarsening stages overlap very strongly, as Figs. 11, 12 and 17
show, and the simple model (\ref{W})-(\ref{g-t_r}) implying the type
of evolution shown in Fig. 3 can hardly be adequate.  Therefore, for
the states $F$ and $G$, only KMC simulations are presented in the
present work. Further refinements of the SSM are evidently needed to
employ this method at the low temperatures or low supersaturations
mentioned.

\section{RESULTS OF SIMULATIONS AND COMPARISON WITH EXPERIMENTS}

\begin{figure}
\caption {(color online \cite{arXiv}). Evolution of the density of
precipitates, $d_p(t)$   (top figure), and their mean radius $R(t)$
(bottom figure) obtained in our simulations for the state $A$ with
$x$=1.17 (in at.\%, here and below), $T$=773 K  for the critical
size $N_c=15$ copper atoms. Right scale shows the number of
precipitates within simulation volume $V_s=(64a)^3$. Solid curves
correspond to the KMCM, and dashed curves, to the SSM. Points show
experimental data by Kolli et al. \cite{KWZS-08} for the
multicomponent steel NUCu-140 with the same $x_{\rm Cu}$=1.17 and
$T$=773 K for $N_c=11$.\label{d_p,R_p-A}}
\end{figure}

\begin{figure}
\caption {(color online \cite{arXiv}). The same as in Fig. 5 but for
the state $B$ with $x$=1.34, $T=773$ K for $N_c=15$. Two dashed
curves (red and green on-line) correspond to the two SSM simulations
with different random number sets. Symbols correspond to
experimental data for the state $B$: circles, to \cite{Goodman-73};
triangles, to \cite{Kampmann-86}; and rhombus, to \cite{Mathon-97}.
\label{d_p,R_p-B}}
\end{figure}

The results of our simulations together with the available
experimental data for some Fe-Cu and Fe-Cu-based alloys are
presented in Figs. 5-17. Figs. 5-12 show the temporal evolution of
the density and the mean radius of precipitates, $d_p(t)$ and
$R(t)$. Note that for the state $F$ in Fig. 11, this evolution is
described in terms of the scaled time $t_s$ defined by Eq.
(\ref{t_s-t_r}) (with replacing $t_r\to t$), as the equilibrium
vacancy concentration $c_v^{eq}(T)$ at the low $T$=561 K considered
is negligibly small, while actually (in particular, in experiments
\cite{Miller-03}) the precipitation at such low $T$ occurs only in
irradiated materials where the vacancies (together with the
interstitial atoms) are formed due to irradiation.

Figs. \hbox{13-16} illustrate the concentration and temperature
dependencies of the maximum precipitate density $d_p^{max}=d_{max}$,
as well as of some temporal characteristics of precipitation,
$t^{max}, \ t^c$ and $t^{c,0.1}$, defined by the relations:
\begin{eqnarray}
&&d_p(t^{max})=d_{max},\label{t^max}\\
&& t^c=t(t_{r3}),\label{t^c}\\
&&d_p(t^{c,0.1})=0.1\,d_{max}\,\label{t^c,0.1}
\end{eqnarray}
where the reduced time $t_{r3}$ and the function $t(t_r)$ in
(\ref{t^c}) are the same as in Eqs. (\ref{t_ri})  and
(\ref{t-t_r_D_i}). The time $t^{max}$ usually corresponds to the
completion of nucleation or beginning of growth; the time $t^c$
approximately corresponds to the onset of coarsening, and
$t^{c,0.1}$ can characterize  the time of completion of the first
stage of coarsening and beginning of its more advanced stages. For
brevity, $t^{max}$ will be called the ``nucleation time'', and
$t^{c,0.1}$, the ``advanced coarsening time''. In Figs. 13-16 we
also present the experimental estimates of $t^{max}$ and $t^{c,0.1}$
but not $t^c$ as such estimates are usually not certain  for the
onset of coarsening.

Fig. 17 illustrates temporal evolution of the precipitate size
distributions observed in our simulations. In caption to this figure
we use the times $t^{N,\alpha}$ and $t^{c,\alpha}$ defined similarly
to the ``nucleation time'' $t^{max}$ and the ``advanced coarsening
time'' $t^{c,0.1}$ in Eqs. (\ref{t^max}) and (\ref{t^c,0.1}):
\begin{eqnarray}
&&d_p(t^{N,\alpha})=\alpha d_{max},\qquad t^{N,\alpha}<t^{max},\label{t^N,alpha}\\
&&d_p(t^{c,\alpha})=\alpha d_{max},\qquad
t^{c,\alpha}>t^{max}\label{t^c,alpha}
\end{eqnarray}
where the number $\alpha$ is less than unity. Times $t^{N,\alpha}$
qualitatively correspond to the nucleation stage, and
$t^{c,\alpha}$, to the coarsening stage.

Let us discuss the results presented in Figs. 5-17. First, Figs.
5-12 and 17 show that the above-mentioned strong attraction between
a vacancy and a copper atom leads to a great difference in the
precipitation kinetics between iron-copper alloys and alloys with no
such attraction, such as the alloys described by simplified models
with time-independent effective direct exchange rates
$\gamma_{\alpha h}^{eff}$ in (\ref{t-t_r}) for which this kinetics
is illustrated by Fig. 3, or the models with a relatively weak
vacancy-minority atom interaction for which this kinetics is
illustrated by Fig. 1 in \cite{SM-00} or Fig. 4 in
\cite{Barashev-04}. For such simplified models, the presence of five
well-defined stages of evolution  shown in Fig. 3 is characteristic,
including the distinct pre-coarsening stage mentioned above. On the
contrary, in Figs. 5-12 such an intermediate pre-coarsening stage is
not seen, the ``pure nucleation'' and ``pure growth''  stages are
relatively short (if exist at all), and the  nucleation, growth and
pre-coarsening stages significantly overlap each other. It is also
illustrated by the size distribution functions $\nu_k(t)$  in Fig.
17  which seem to imply that the nucleation, growth and
pre-coarsening processes at $0.5\,t^{max} \lesssim t\lesssim
t^{max}$  occur simultaneously. As discussed in \cite{KSSV-11},
these kinetic features seem to be mainly related to the strong
vacancy trapping by copper precipitates, which leads to a great
acceleration of both the growth and the Ostwald ripening processes
as compared to the alloys with no such trapping.

Figs. 5-10 also show that the simplified direct-atomic-exchange
model (\ref{SKE})-(\ref{F_c-BCC}) using the above-mentioned maximum
thermodynamic gain principle and the simple interpolation
(\ref{t-t_r_D_i}) describes the precipitation kinetics in a fair
agreement with the KMCM. Discrepancies between the SSM and the KMCM
results lie usually within statistical errors of simulations, and
these discrepancies can be still more reduced if the smoothed
interpolations mentioned above are used instead of more simple
expressions (\ref{t-t_r_D_i}).

Figs. 5, 6 and 8-10 also illustrate opportunities to use the SSM to
extrapolate the KMC simulations to the longer evolution times. As
mentioned, the KMC simulations of coarsening are time-consuming,
unlike the SSM simulations. At the same time, the above-described
considerations about the physical nature of the ``rescaling of
time'' function $t(t_r)$ in (\ref{t-t_r_D_i}) enable us to expect
that this function preserves its form  for the first stages of
coarsening at least up to $t\sim t^{c,0.1}$, until the later, more
advanced stages of coarsening (including the asymptotic
Lifshits-Slyozov-Wagner stage \cite{Lif-61,Wag-61}) start.
Therefore, we can use the SSM simulated $d_p(t_r)$, $R(t_r)$ and Eq.
(\ref{t-t_r_D_i}) to estimate the $d_p(t)$ and $R(t)$ values for
those $t$ for which the KMC results are not available. This is shown
in Figs. 5, 6 and 8-10.

\begin{figure}
\caption {(color online \cite{arXiv}). The same as in Fig. 6 but for
seven different KMC simulations for $N_c=10$. Thick solid line (red
on-line) corresponds to the same KMC simulation as that shown in
Fig. 6 by thick line for the choice $N_c=15$.\label{d_p,R_p-B-10}}
\end{figure}

\begin{figure}
\caption {(color online \cite{arXiv}). The same as in Fig. 5 but for
the state $C$ with $x$=1.82, $T=773$ K for $N_c=12$. Points show
experimental data by Kolli et al. \cite{KS-08,KWZS-08} for the
multicomponent steel NUCu-170 with the same $x_{\rm Cu}=1.82$,
$T=773$ K for $N_c=11$.\label{d_p,R_p-C}}
\end{figure}

Figs. 6 and 7 also illustrate some methodical points.  Two dashed
curves in Fig. 6 (red and green on-line) correspond to two SSM
simulations with the different random number sets, thus their
difference illustrates the statistical scatter of the SSM simulation
results. We see that this scatter is significant only at the very
end of simulations when the total precipitate number $N_p$ becomes
small. Similarly, seven different KMC simulations shown in Fig. 7
illustrate the statistical scatter of the KMC simulation results.
Influence of the choice of the ``critical'' size $N_c$ in Eqs.
(\ref{d_p-R_p-def})  on the simulated $d_p(t)$ and $R(t)$ is
illustrated by Figs. 6 and 7 which correspond to the same alloy
state $B$ but to the different $N_c$, 15 and 11 copper atoms,
respectively, while the thick solid curve in both Fig. 6 and Fig. 7
(red on-line in Fig. 7) corresponds to the same KMC simulation. We
see that the variations of $N_c$ used make noticeable effects on the
simulated $d_p(t)$ and $R(t)$ only at the first stages of
nucleation, while later on  such effects become insignificant.
\begin{figure}
\caption {(color online \cite{arXiv}). The same as in Fig. 5 but for
the state $D$ with $x$=1.34, $T=713$ K for $N_c=11$.
\label{d_p,R_p-D}}
\end{figure}
\begin{figure}
\caption {(color online \cite{arXiv}). The same as in Fig. 5 but for
the state $E$ with $x$=1.34, $T=663$ K for $N_c=8$.
\label{d_p,R_p-E}}
\end{figure}

Let us discuss the concentration and temperature dependencies of the
kinetic characteristics of precipitation presented in Figs. 13-16.
For the maximum precipitate density $d_{max}$, these dependencies
shown in Figs. 13 and 15 are mainly determined by the reduced
supersaturation $s$ (defined in Eq. (\ref{s}) and Table 1) which
characterizes the scale of the thermodynamic driving force for
precipitation. For the given temperature $T$ or the given
concentration $x$, the  $d_{max}$ value increases with $s$, and at
low $s\lesssim 0.3$ or high $T\gtrsim 800$ K this rise is rather
sharp. At the same time, Fig. 15 shows that at not low $s\gtrsim
0.3-0.4$ and not high $T$, the $d_{max}$ value changes with $T$ more
slowly, and at $T\lesssim 700$ K, the further lowering temperature
makes little effect on the $d_{max}$. It may imply that the slowing
down of kinetics due to the strong vacancy-copper atom correlations
discussed below becomes important for these $T$. The temporal
characteristics of precipitation shown in Figs. 14 and 16$b$
decrease with increasing supersaturation $s$, which can be explained
by an increase of the thermodynamic driving force. At the same time,
Fig. 16$a$ shows that  at not low $s\gtrsim 0.3-0.4$ and not high
$T\lesssim 750$ K, the scaled times $t_s^{max}$, \ $t_s^c$ and
$t_s^{c,10}$ vary with temperature rather weakly. It can be
explained by an interplay between an increase with lowering $T$ of
both the thermodynamic driving forces  which promote the evolution,
and the vacancy-copper atom correlations which reduce the copper
diffusivity $D_{\rm Cu}$ \cite{SF-07} and thereby slow down the
evolution.

Let us discuss the precipitate size distributions $\nu_k(t)$ shown
in Fig. 17. First, we see that these distributions are usually
rather broad, and sizes of different precipitates are typically very
different. Therefore, characterization of these sizes by only their
mean value $R(t)$ used in Figs. 5-12 is oversimplified and
incomplete. It is true not only for the coarsening stage (for which
a great difference in the precipitate sizes is natural as the bigger
precipitates coarsen due to dissolving the smaller ones) but also
for  all other stages of precipitation. It can be related to the
strong overlapping of the nucleation, growth and pre-coarsening
stages mentioned above. Comparison of the size distributions
$\nu_k(t)$ at $t$=$t^{max}$ and $t$=$1.1\,t^{max}$ (that is, frames
G2 and G3, B2 and B3, or  D2 and D3 in Fig. 17) also illustrates the
significant overlapping of growth and  early coarsening stages. We
see that the great majority of precipitates at $t$=$1.1\,t^{max}$
still continue to grow due to the absorption of copper atoms from
the matrix, even though some smallest precipitates already start to
dissolve. Such significant overlapping of growth and  early
coarsening stages agrees with the observations by Mathon et al. for
the Fe-1.34Cu alloys (the state $B$ in Table 1) \cite{Mathon-97},
and it was also noted by Kolli et al. for the NUCu steels
\cite{KWZS-08}. Second, the comparison of the size distributions
$\nu_k(t)$ at $t$=$t^{max}$ and $t$=$1.1\,t^{max}$ also illustrates
very sharp variations of these distributions with the evolution time
$t$, in particular, at $t$ corresponding to the beginning of
coarsening, which is not clearly seen in the ``averaged''
description of Figs. 5-12. Third, Fig. 17 shows that the precipitate
size distribution $\nu_k(t)$ for the state $G$ with the highest
temperature $T$=873 K is, generally, much more uniform than
$\nu_k(t)$  for the states $B$ and $D$ with the lower $T$=773 K and
713 K, particularly for the early coarsening stage, which is
illustrated by frames G4, B4 and D4. In particular, in frames B4 and
D4 we observe only one very big cluster with $k$$\sim$600 copper
atoms, numerous small clusters with $k$$<$200 copper atoms, and 3-4
``middle-sized'' clusters with $k$$\sim$200-300 atoms (the same
features are also observed in $\nu_k(t)$ for the state $A$ not shown
in Fig. 17 on considerations of space), in the great difference with
frame G4 (as well as G2 and G3) where the cluster size distribution
is rather uniform. This difference can be related to the weakening
of the vacancy-copper atom correlations at high $T$ which can
enhance the copper diffusivity $D_{\rm Cu}$ and thereby promote
growth of many big precipitates for the state $G$ unlike the states
$A$, $B$ and $D$, but these points need further studies. Finally,
the lower row of Fig. 17 illustrates features of precipitation at
low temperatures $T\sim$ 300$^{\rm o}$ C when the vacancies
(necessary for the atomic diffusion) are provided by irradiation.
Figs. F1--F3 illustrate a very strong overlapping of the nucleation
growth and pre-coarsening stages, while the coarsening stage was not
reached in these our simulations.

Let us now compare the simulation results to the available
experimental data. For the state $B$ and the KMC simulations shown
in Figs. 6 and 7 by thick lines, the detailed comparison with
various experiments was given by Soisson and Fu \cite{SF-07} who
concluded that the predictions of simulations are reliable. The
results presented in Figs. 6, 7, 14 and 16 can complement their
discussion by the two points. First, Figs. 6 and 7 confirm that the
disagreements between simulations and experimental observations seem
usually to lie within statistical errors of simulations and the
scatter of experimental results. Second, Figs. 6, 14 and 16 show
that the SSM-based extrapolations of KMC simulations for the first
stages of  coarsening seem to agree with the observations as well.

\begin{figure}
\caption {The same as in Fig. 5 but for the state $F$ $x$=0.78,
$T=561$ K for $N_c=5$. \label{d_p,R_p-F}}
\end{figure}
\begin{figure}
\caption {The same as in Fig. 5 but for the state $G$ with $x$=1.15,
$T=873$ K for $N_c=28$. Solid symbols show experimental data by
Shabadi et al. \cite{Shabadi-11} for the following alloys aged at
$T=873$ K: circles, binary Fe-1.15Cu; squares, ternary
Fe-1.14Cu-0.99Mn. Right scale shows the number of precipitates,
$N_p$,  within simulation volume $V_s=(128a)^3$. \label{d_p,R_p-G}}
\end{figure}

For the state $G$ with $x=1.15$, $T=873$ K,  our simulations are
compared to the data by Shabadi et al. \cite{Shabadi-11} in Fig. 12.
As supersaturation $s$ and the precipitate density $d_p$ for this
state are rather low, these simulations are time-consuming and
include only nucleation and growth stages, while the data by Shabadi
et al. seem to correspond to the longer aging times and have
significant errors. Within these errors, the simulation and
experimental results in Fig. 12 can be considered as agreeing with
each other, particularly for the maximum precipitate density
$d_{max}$, though the simulated evolution times can be somewhat
shorter than the observed ones. Solid squares in Fig. 12 correspond
to a ternary Fe-Cu-Mn alloy and illustrate the effect of the third
alloying element Mn on the precipitation kinetics. The presence of
Mn seems to lead to an increase of the maximum precipitate density
$d_{max}$ by about twice with respect to the analogous Fe-Cu binary,
in a qualitative contrast with the effect of alloying elements  on
the $d_{max}$ value in the NUCu steels discussed below.
\begin{figure}
\caption {Concentration dependence of the maximum density of
precipitates, $d_{max}$, for decomposition of Fe-$x$Cu and
Fe-Cu-based alloys at $T=773$ K. Open symbols here and below
correspond to our simulations, and solid symbols, to experiments.
Solid circle corresponds to experiments \cite{Kampmann-86} for the
state $B$, while solid square and solid triangle, to experiments
\cite{KS-08,KWZS-08} for the multicomponent steel NUCu-140 with
$x_{\rm Cu}$=1.17 and NUCu-170 with $x_{\rm Cu}$=1.82, respectively.
Dashed lines here and below are given to guide the eye.
\label{d_max-c}}
\end{figure}
\begin{figure}
\caption {Concentration dependence  of temporal characteristics of
precipitation: $t^{max}$ (circles), $t^c$ (triangles), and
$t^{c,0.1}$ (squares), defined in the text by Eqs. (\ref{t^max}),
(\ref{t^c}) and  (\ref{t^c,0.1}), respectively, at $T=773$ K. Solid
circle and solid square at $x_{\rm Cu}$=1.17 or $x_{\rm Cu}$=1.82
correspond to the data \cite{KS-08,KWZS-08} for the steel NUCu-140
or NUCu-170, respectively, while solid circle and solid square at
$x$=1.34 correspond to the data \cite{Goodman-73}-\cite{Mathon-97}
for the state B. \label{t^max-t^c-c}}
\end{figure}
\begin{figure}
\caption {Temperature dependence of the maximum density of
precipitates in the course of decomposition of  Fe-$x$Cu and
Fe-Cu-based alloys. Open circles correspond to the states $B$, $D$
and $E$ with $x$=1.34; open square, to  the state $A$ with $x$=1.17;
and open triangle, to the state $G$ with $x$=1.15. Solid circle
corresponds to experiments \cite{Kampmann-86} for the state $B$ with
$x$=1.34; solid square, to experiments \cite{KWZS-08} for the steel
NUCu-140 with $x_{\rm Cu}$=1.17; and solid triangle, to experiments
\cite{Shabadi-11} for the state $G$ with $x$=1.15. \label{d_max-T}}
\end{figure}

In Figs. 5, 8, and 13-16 we compare the simulation results for the
state $A$ or $C$  with  $T=773$ K  and $x=1.17$  or 1.82 to the data
by Kolli et al. \cite{KS-08,KWZS-08} for the multicomponent steels
NUCu-140 and NUCu-170 with the same $T$ and $x_{\rm Cu}$. In
addition to copper, these steels contain a number of alloying
elements: C, Al, Ni, Si, Mn, Nb, P and S, 5.49\% on the total in
NUCu-140, and 5.83\%, in NUCu-170, while the partial concentrations
of each alloying element in these two steels are very close to each
other \cite{KS-08,KWZS-08}. Therefore, differences in the
precipitation kinetics for these two steels can be mainly related to
the difference in the copper content $x_{\rm Cu}$. Then comparison
of this kinetics for each of these steels to that for the analogous
binary Fe-$x$Cu alloy can elucidate the effect of alloying elements
on the precipitation at different $x_{\rm Cu}$. Qualitatively, these
problems were discussed by Kolli et al. \cite{KWZS-08}. Our
simulations enable us to consider these points quantitatively. Let
us also note that the critical sizes $N_c$ for these two steels
estimated by Kolli et al. \cite{KS-08,KWZS-08}: $N_c\simeq$11 copper
atoms, are rather close to the estimates for their binary analogues
presented in Table 1: $N_c^{th}(A)\simeq 15$, $N_c^{th}(C)\simeq 12$
copper atoms.

Let us first discuss the nucleation and growth stages illustrated by
Figs. 5 and 8. For the NUCu-140 or NUCu-170 steel this corresponds
to $t\lesssim 1$ h and $t\lesssim 0.25$ h, respectively, and the
maximum precipitate density $d_{max}$ in each steel is lower than
that in its binary analogue by about three times. However, for the
NUCu-140, both the values and the temporal dependencies of $d_p(t)$
and $R(t)$ in Fig. 5 seem to not greatly differ from those simulated
for the Fe-1.17Cu alloy, particularly for the nucleation stage, and
the nucleation time $t^{max}$ can be similar, too. On the contrary,
for the NUCu-170, the data at $t$=0.25 h shown in Fig. 8,
particularly for the $R(t)$ value, sharply disagree with those
simulated for the Fe-1.82Cu alloy, while the nucleation time
$t^{max}$ exceeds that for the Fe-1.82Cu alloy by an order of
magnitude.

Therefore, our comparison  seems to imply that the effect  of almost
the same content of alloying elements on  the nucleation kinetics in
NUCu-170 with the higher copper content $x_{\rm Cu}$=1.82 is much
stronger and qualitatively different from that  in NUCu-140 with the
lower $x_{\rm Cu}$=1.17. Physically, such conclusion does not seem
to be natural. In this connection we note that this conclusion is
mainly based on the data for the mean precipitate size in NUCu-170
at $t$=0.25 h  reported  by Kolli and Seidman \cite{KS-08}:
\hbox{$R\simeq 1.2$ nm.} This value much exceeds the critical radius
$R_c\simeq 0.3$ nm estimated for this steel in \cite{KS-08}:
$R\simeq 4R_c$. It should imply that in the course of the nucleation
stage (supposed in \cite{KS-08} for the NUCu-170 at $t$=0.25 h to
explain a steep rise of the precipitate density $d_p(t)$ between
$t$=0.25 h and $t$=1 h seen in Fig. 8) the new-born precipitates
grow extremely fast. Such a very sharp growth at the early
nucleation stage seems to be very unusual and, to our knowledge, was
never observed in either experiments or simulations, as illustrated
by Figs. 5-12. Therefore, the data about $R(t)$ in NUCu-170 at
$t$=0.25 h reported in \cite{KS-08} should possibly be taken with
some caution.

\begin{figure}
\caption {Temperature dependence of the same temporal
characteristics of precipitation  as those in Fig. 14, but in terms
of the scaled time $t_s$ defined by Eq. (\ref{t_s-t_r}): $t_s^{max}$
(circles), $t_s^c$ (triangles), and $t_s^{c,0.1}$ (squares). In
figure $a$,  the states $B$, $D$ and $E$ have the same $x$=1.34,
while solid circle and solid square correspond to experiments
\cite{Kampmann-86} for the state $B$ with  $x$=1.34. In figure $b$,
the state $A$ corresponds to $x$=1.17; solid circle and solid square
at $T$=773 K correspond to the data \cite{KWZS-08} for the steel
NUCu-140 with $x_{\rm Cu}$=1.17; and solid circle at $T$=873 K
corresponds to experiments \cite{Shabadi-11} for the state $G$ with
$x$=1.15. \label{t^max-t^c-T}}
\end{figure}

\begin{figure}
\caption {Numbers of clusters containing $k$ copper atoms,
$\nu_k(t)$,  observed in our KMC simulations. The first, second,
third and fourth row corresponds to the state $G$, $B$, $D$ and $F$
in Table 1, respectively. Frame G1, G2, G3, or G4 corresponds to the
time $t$ equal to  $t^{N,0.47}$, $t^{max}$,
$1.1\,t^{max}$=$t^{c,0.87}$, or $t^{c,0.67}$; frame B1, B2, B3, or
B4, to $t^{N,0.61}$, $t^{max}$, $1.1\,t^{max}$=$t^{c,0.94}$, or
$t^{c,0.61}$; and frame D1, D2, D3, or D4, to $t^{N,0.6}$,
$t^{max}$, $1.1\,t^{max}$=$t^{c,0.86}$, or $t^{c,0.61}$,
respectively, where the time $t^{N,\alpha}$, $t^{max}$  or
$t^{c,\alpha}$ is defined by Eq. (\ref{t^N,alpha}), (\ref{t^max}),
or (\ref{t^c,alpha}). Frame F1, F2, or F3 corresponds to the scaled
time $t_s$ equal to $t_s^{N,0.48}$, $t_s^{N,0.94}$, or $t_s^{max}$,
where the time $t_s^{N,\alpha}$ is defined by Eq. (\ref{t^N,alpha})
with replacing $t$$\to$$t_s$, while $t_s^{max}=1.45$ h corresponds
to the end of our simulations for the state $F$. \label{nu_k(t)}}
\end{figure}

For the coarsening stage, the results presented in Figs. 5, 8 and 14
fully agree with the main conclusions of Kolli et al. \cite{KWZS-08}
about a very strong slowing down of coarsening in the NUCu steels
with respect to binaries Fe-Cu. In particular, the advanced
coarsening time $t^{c,0.1}$ for each of these steels exceeds that
for its binary analogue by about two orders of magnitude.  At the
same time, the dependencies of this  advanced coarsening time
$t^{c,0.1}$ on the copper content $x_{\rm Cu}$  for the NUCu steels
and for their binary analogues shown in Fig. 14 seem to be similar.

The strong slowing down of coarsening in a ternary Fe-Cu-Mn alloy
with respect to its binary analogue Fe-Cu was also observed by
Miller et al.  under neutron irradiation \cite{Miller-03}. On the
contrary,  the effects of alloying elements on the nucleation and
growth  kinetics in Fe-Cu-Mn alloys and in NUCu steels seem to
differ qualitatively: according to Figs. 12 and 13, the $d_{max}$
value for an Fe-Cu-Mn alloy  is by about twice higher, while for
each of NUCu steels, it is by about three times lower than that in
its binary analogue. Therefore, for the coarsening stage, the
effects of alloying elements on the decomposition kinetics in the
multicomponent Fe-Cu-based alloys seem to be much more universal
than those for the earlier stages of precipitation.

Such a universal  slowing down of coarsening in the multicomponent
Fe-Cu-based alloys with respect to their binary analogues Fe-Cu can
be related to a  significant segregation of alloying elements on the
surface of precipitates \cite{Miller-03}-\cite{KWZS-08} which can
reduce the surface energy and thereby the thermodynamic driving
force for coarsening. It can also be related to a weakening of the
vacancy trapping at surfaces of precipitates due to this
segregation. However, quantitative estimates of these effects seem
to be absent yet.

\section{Conclusions}

Let us summarize the main results of this work. The
earlier-developed $ab$ $initio$ model and both the kinetic Monte
Carlo (KMC) and the stochastic statistical methods are used to
simulate the precipitation kinetics for seven binary Fe-Cu alloys
with different copper concentrations $x$ and temperatures $T$.
Comparison of results obtained to available experimental data and to
other simulations enable us to make a number of conclusions about
kinetic features of precipitation in both the binary Fe-Cu and the
multicomponent Fe-Cu-based alloys.

First, we find that due to the strong vacancy trapping by copper
precipitates, the precipitation kinetics in iron-copper alloys for
all $x$ and $T$ considered differs notably from that observed for
the alloys with no such trapping: the ``pure nucleation'' and ``pure
growth'' stages are relatively short, the nucleation, growth and
coarsening stages significantly overlap, while the intermediate
``pre-coarsening'' stage observed in some simulations for simplified
alloy models (illustrated by Fig. 3 of this work and  by Fig. 1 in
Ref. \cite{SM-00}) is absent. In this connection, the presence of
this pre-coarsening stage in simulations of precipitation in
irradiated Fe-Cu alloys made by Barashev et al. \cite{Barashev-04}
can be related just to some oversimplifications of their model.

The concentration and temperature dependencies of the maximum
precipitate density $d_{max}$, the nucleation time $t^{max}$, and
the advanced coarsening time $t^{c,0.1}$ defined by Eqs.
(\ref{t^max}) and (\ref{t^c,0.1}) are illustrated by Figs. 13-16. At
low supersaturations $s$, these dependencies are rather sharp and
seem to be mainly determined by the variations of supersaturation
$s(x,T)$ with $x$ or $T$. At higher $s\gtrsim$0.3, these temperature
dependencies become more smooth and seem to be determined by an
interplay between an  enhancement with lowering $T$ of both the
thermodynamic driving forces promoting the evolution and the
vacancy-copper atom correlations reducing the copper diffusivity
$D_{\rm Cu}$ and thereby slowing down the evolution.

Temporal evolution of the precipitate size distributions $\nu_k(t)$
is illustrated by Fig. 17. These distributions are typically rather
broad, and they strongly vary with the evolution time $t$.
Therefore, the conventional description of these sizes in terms of
the mean precipitate size $R(t)$ is oversimplified and incomplete.
We also find that for the alloy state $G$ with a relatively high
temperature $T$=873 K, the size distribution $\nu_k(t)$ is much more
uniform than those observed for the states $B$ and $D$ with the
lower temperatures $T$=773 K and $T$=713 K, particularly for the
coarsening stage.

We also describe an improved version of the earlier-suggested
stochastic statistical method for simulations of precipitation and
show that this version can be used for various extrapolations of KMC
simulations, in particular, for their extensions to the first stages
of coarsening for which the KMC simulations are time-consuming.

Comparison of our simulated temporal dependencies for the density
and the mean size of  precipitates in binary Fe-$x$Cu alloys at
$x$=1.34, $T$=773 K and $x$=1.15, $T$=873 K  to the available
experimental data \cite{Goodman-73,Kampmann-86,Mathon-97,Shabadi-11}
shows a reasonable agreement within both statistical errors of
simulations and the scatter of experimental results. The sizes $N_c$
of critical precipitates calculated by the statistical method of
Dobretsov and Vaks \cite{DV-98} and presented in Table 1 are close
to those estimated in our KMC simulations and in experiments by
Kolli et al. \cite{KS-08,KWZS-08} for the NUCu steels.

Comparison of results of our simulations for the Fe-1.17Cu and
Fe-1.82Cu alloys to the data by Kolli et al. \cite{KS-08,KWZS-08}
about precipitation in NUCu-140 and NUCu-170 steels which have the
same copper content, $x_{\rm Cu}=1.17$ and $x_{\rm Cu}=1.82$, and
contain the similar amount of other alloying elements enables us to
assess the effects of these alloying elements on the precipitation
kinetics. The maximum precipitate density $d_{max}$ in each of these
two steels is lower than that in its binary analogue by about three
times. For the nucleation stage, the precipitate density $d_p(t)$
and their mean size $R(t)$ observed by Kolli et al. \cite{KWZS-08}
in the NUCu-140 steel seem to be close to those simulated for the
Fe-1.17Cu alloy, contrary to the case of the NUCu-170 steel for
which the $d_p(t)$ and $R(t)$ values at $t$=0.25 h reported by Kolli
and Seidman \cite{KS-08} sharply disagree with those simulated for
the Fe-1.82Cu alloy. In this connection we note that the $R(0.25
{\rm h})$=$R_{\rm KS}$ value reported in \cite{KS-08} seems to be
unrealistically large for the early nucleation stage supposed by
Kolli and Seidman for the NUCu-170 steel at $t$=0.25 h: $R_{\rm
KS}\simeq 4R_c$. Therefore, further experimental studies of the
nucleation kinetics in Fe-Cu-based steels seem to be desirable.

For the coarsening stage, the presence of alloying elements in the
NUCu steels leads to a very strong slowing down of coarsening, by
1-2 orders of magnitude, as compared to their binary analogues. A
similar strong slowing down of coarsening was also observed  by
Miller et al. \cite{Miller-03}) for an irradiated ternary Fe-Cu-Mn
alloy. At the same time, for the nucleation and growth stages, the
effects of alloying elements on the maximum precipitate density
$d_{max}$ in the NUCu steels and in the Fe-Cu-Mn alloy studied by
Shabadi et al. \cite{Shabadi-11} seem to be qualitatively different.
Some hypotheses about a possible origin of the universal slowing
down of coarsening in multicomponent Fe-Cu-based alloys  with
respect to their binary analogues are suggested.

 \vskip5mm
We are very grateful to Frederic Soisson for providing to us the
kinetic Monte Carlo codes used for all KMC simulations of this work,
as well as for the valuable critical remarks. We are also much
indebted to Dr. R. Shabadi for sending to us a preprint of paper
\cite{Shabadi-11} prior to publication. The work was supported by
the Russian Fund of Basic Research (grant No. 12-02-00093); by the
fund for support of leading scientific schools of Russia  (grant No.
NS-215.2012.2); and by the program of Russian university scientific
potential development (grant  No. 2.1.1/4540).

\newpage

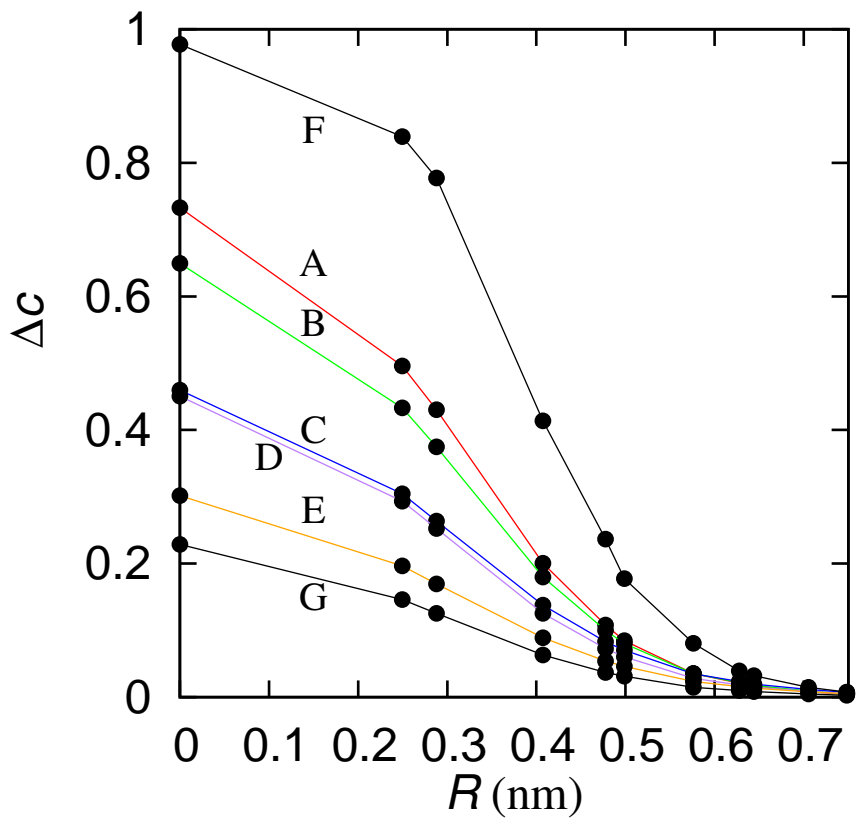
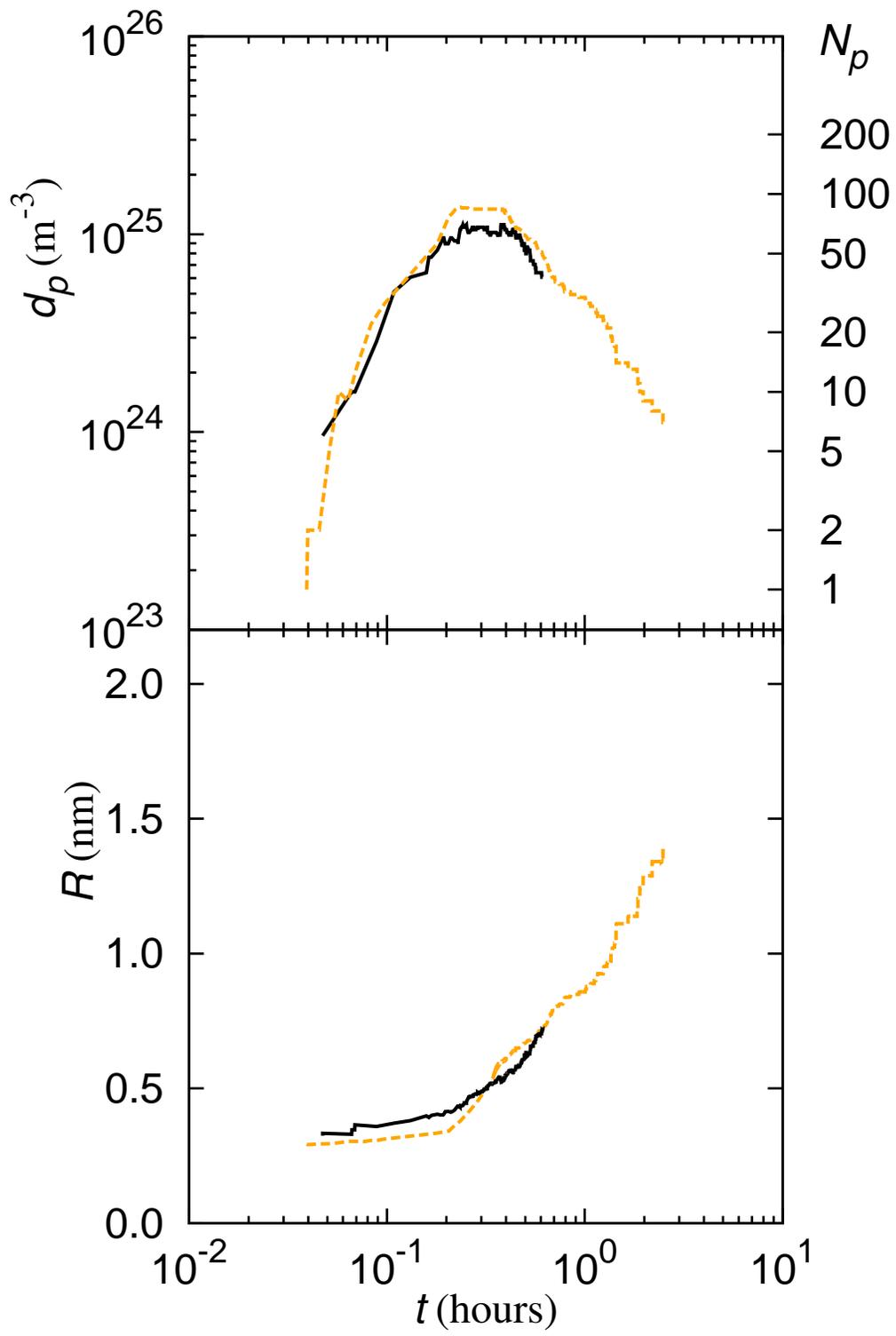
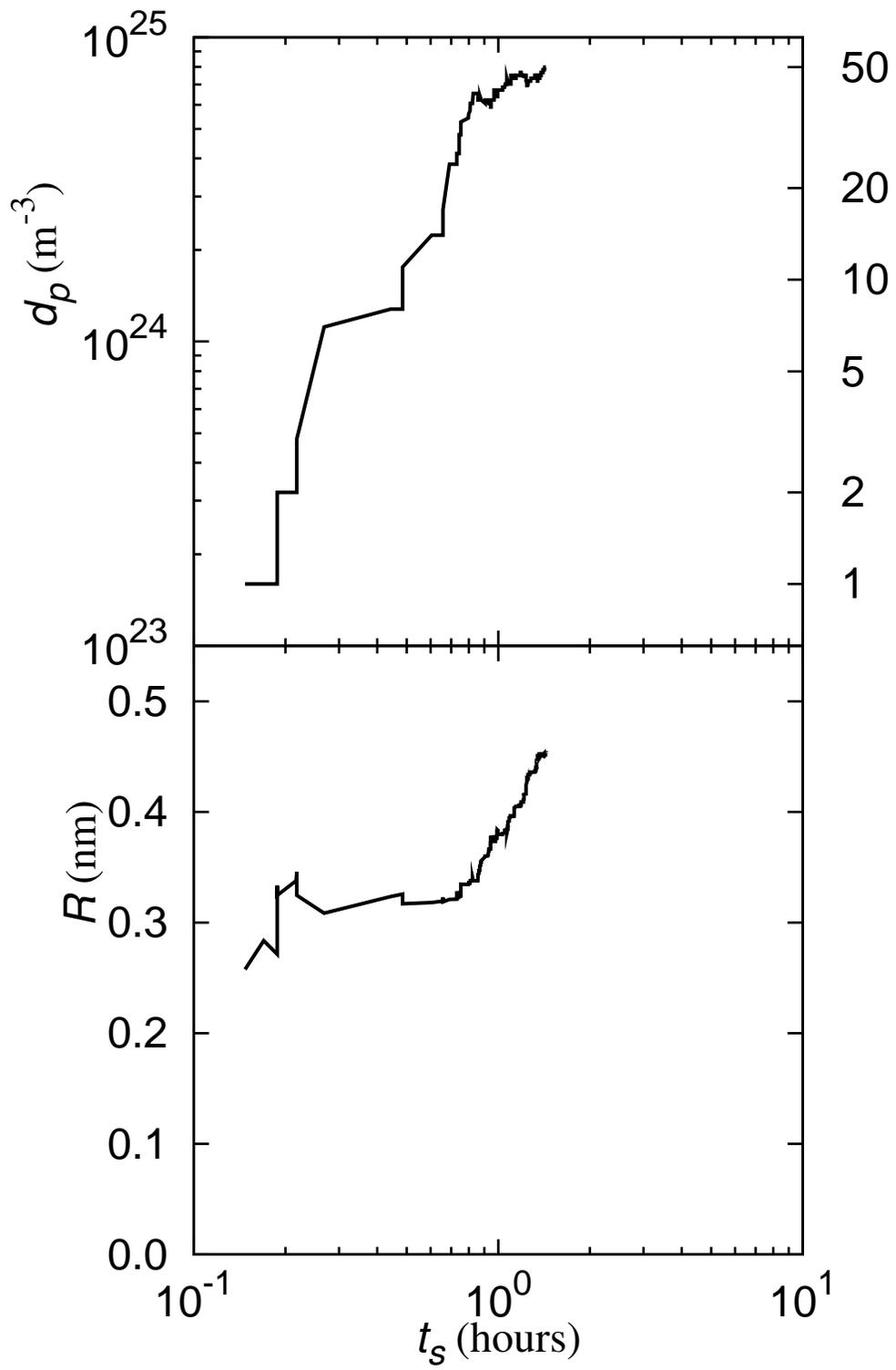
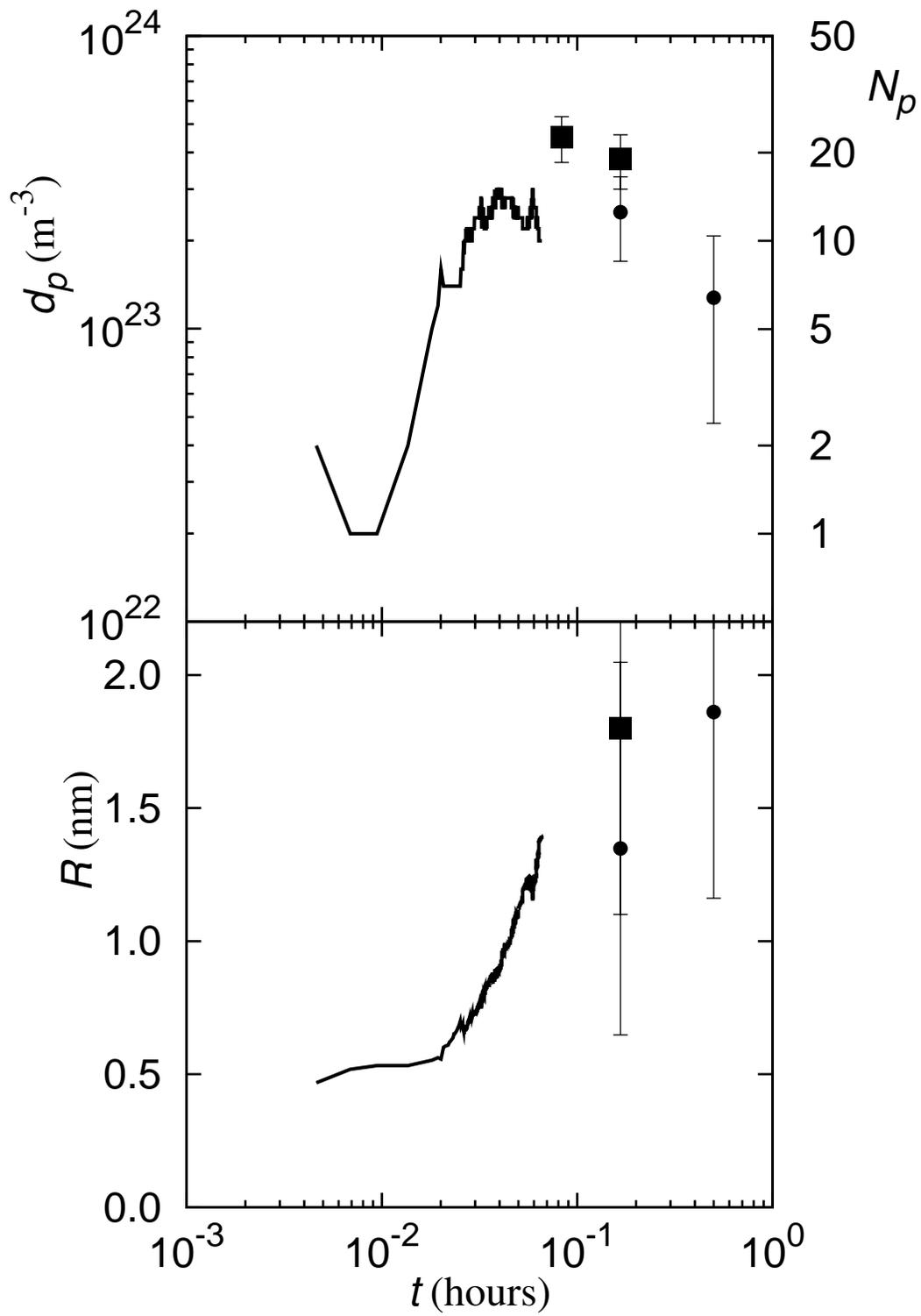
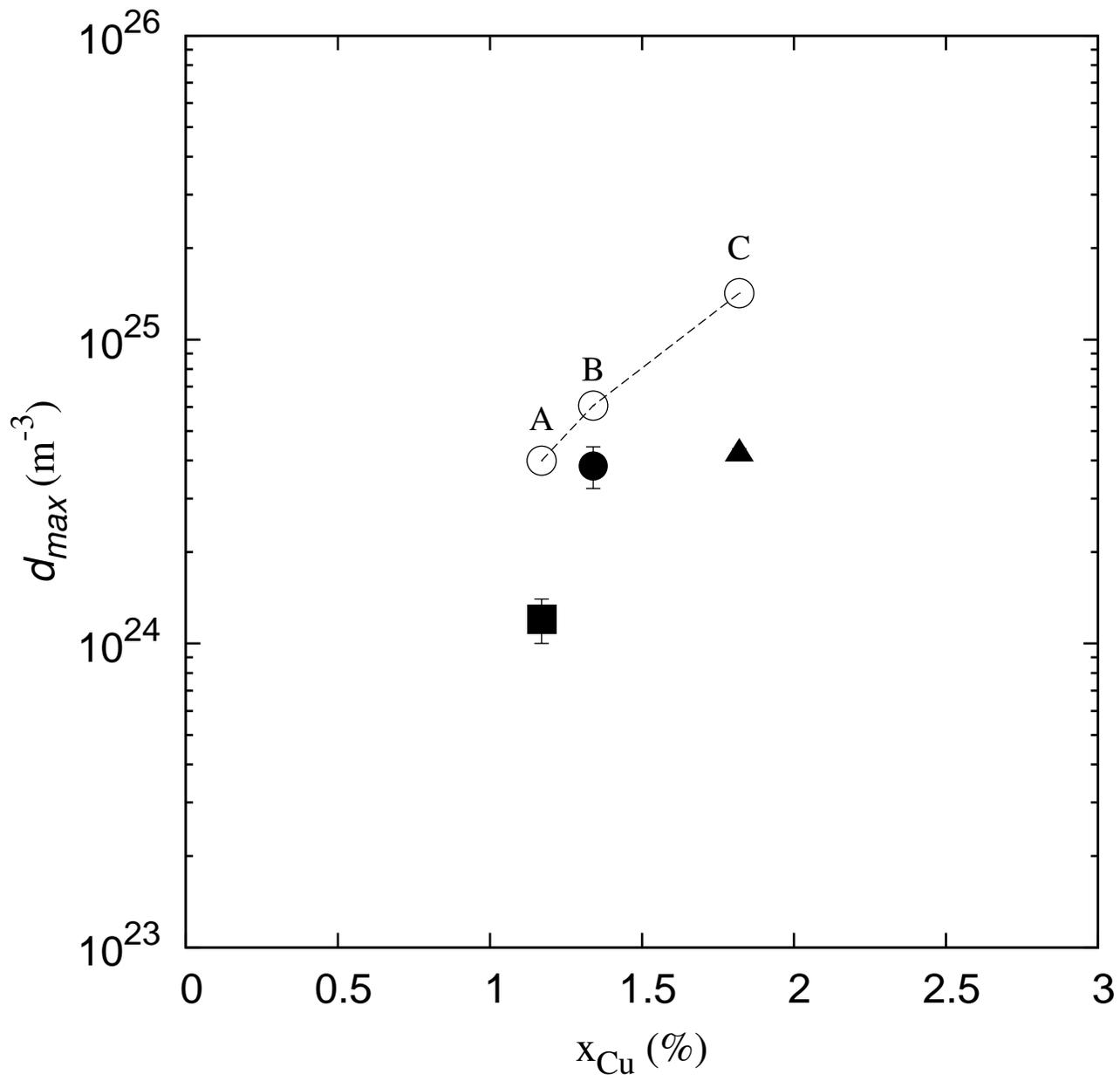
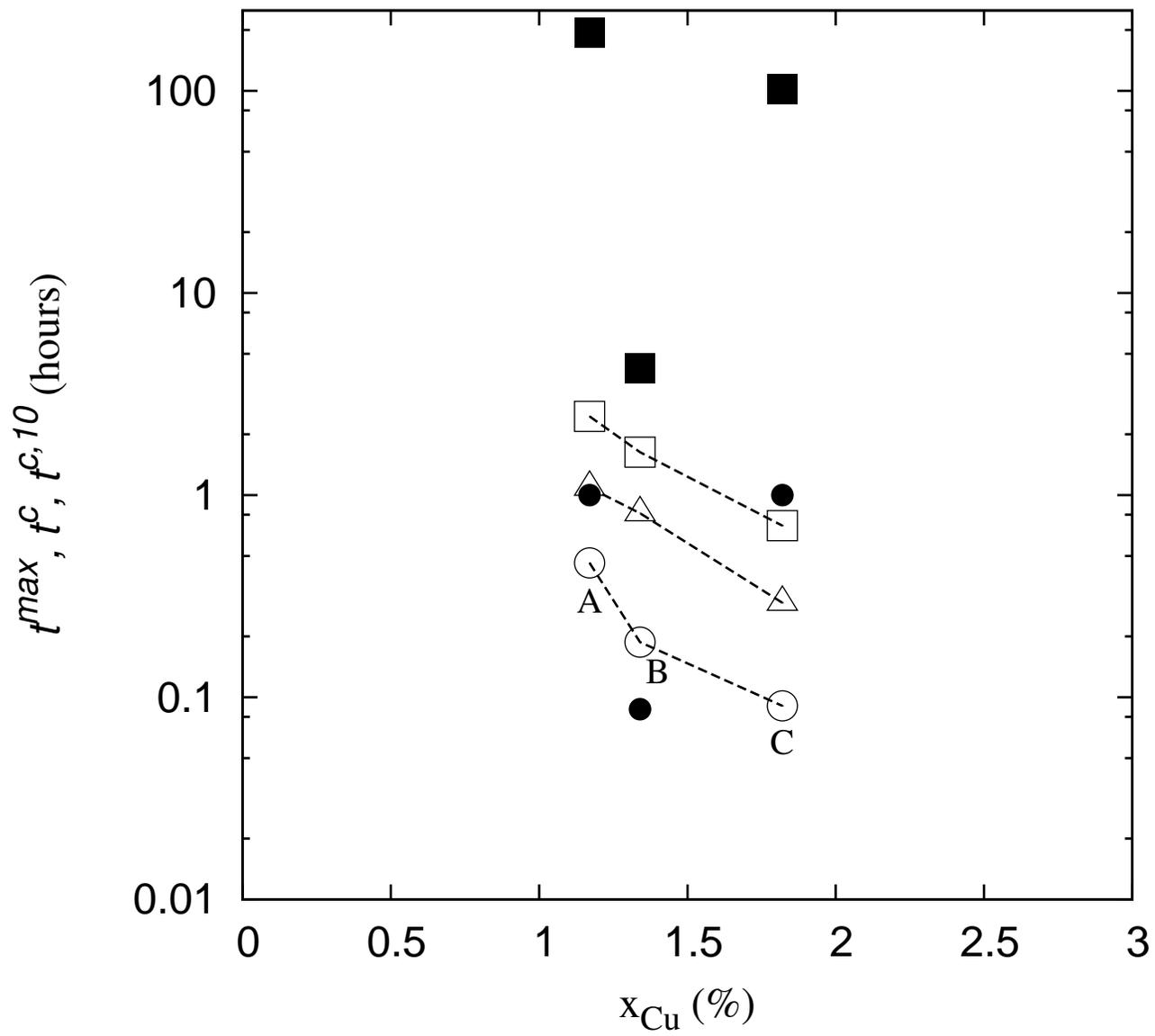
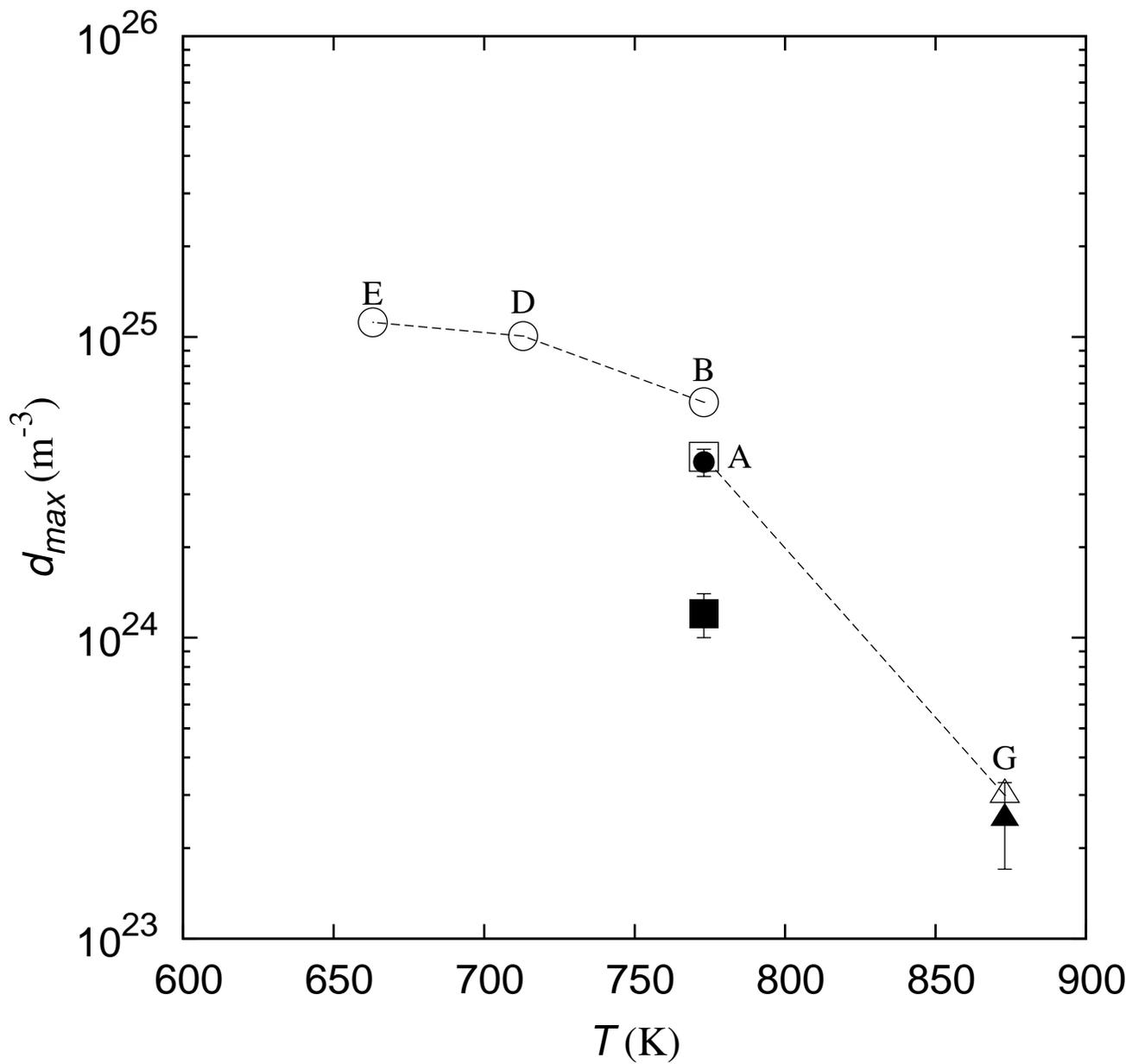
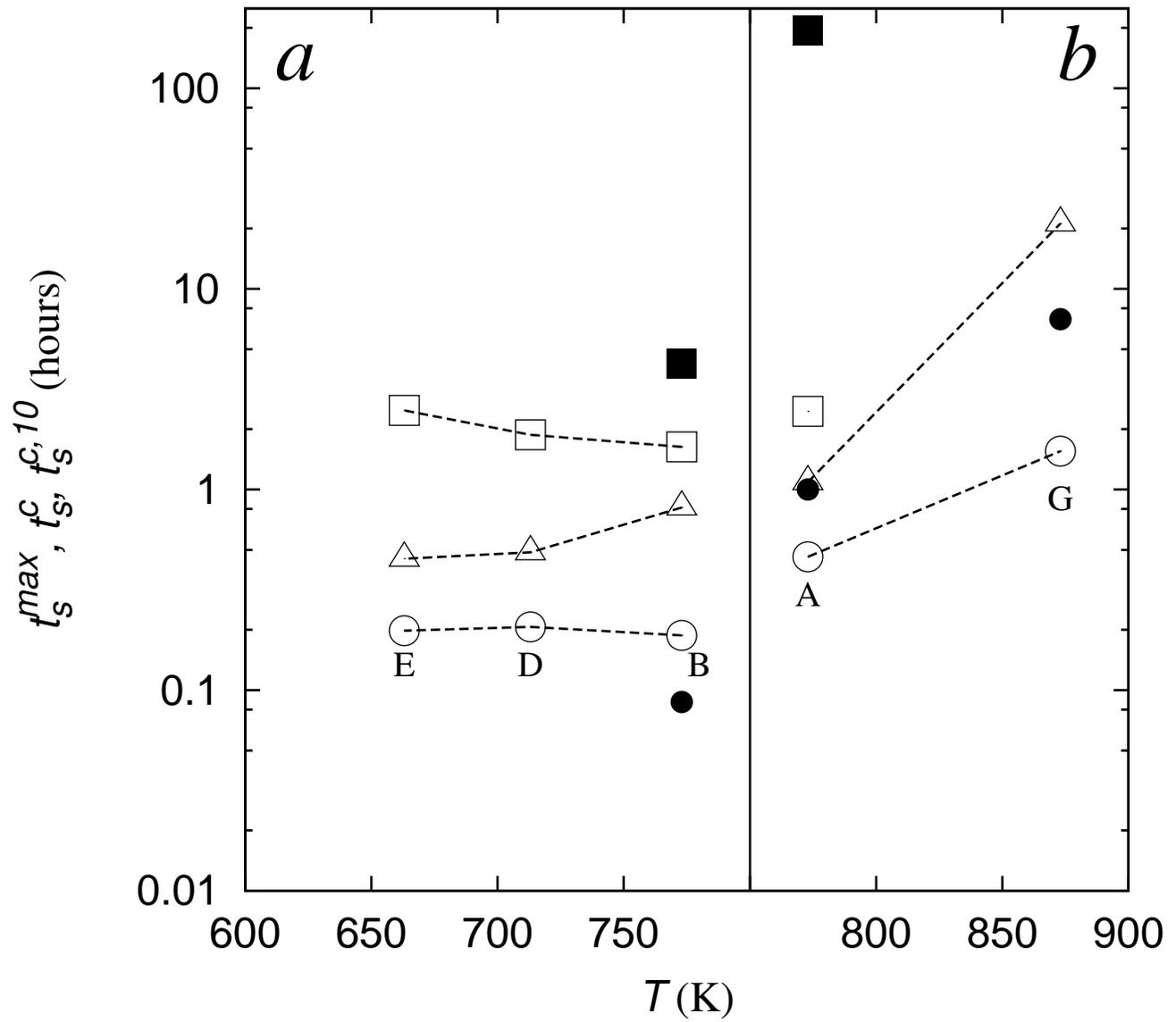
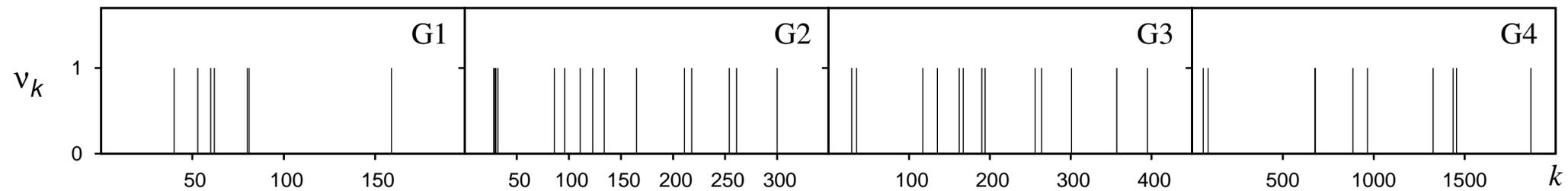
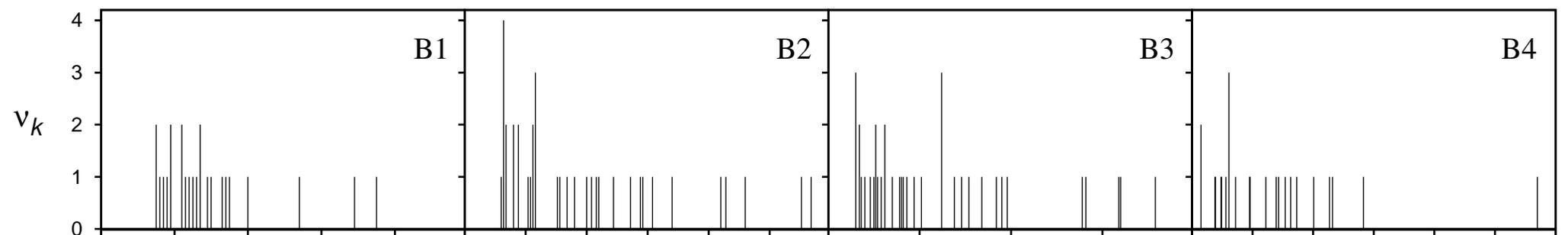
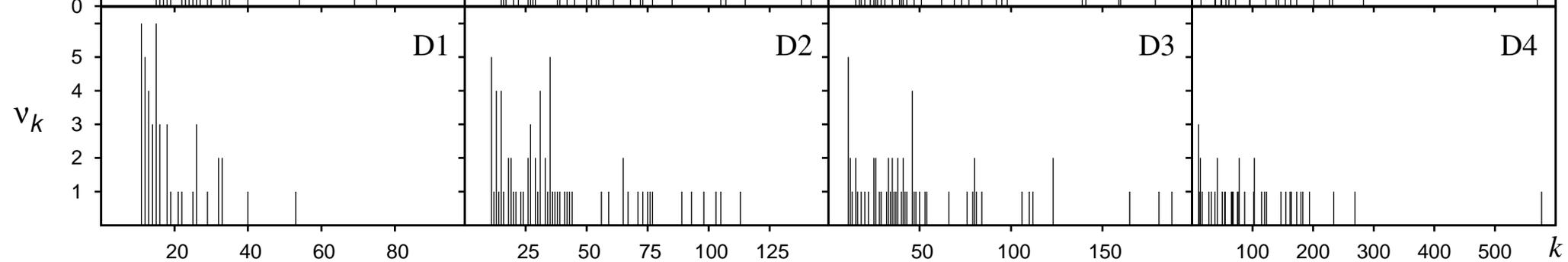
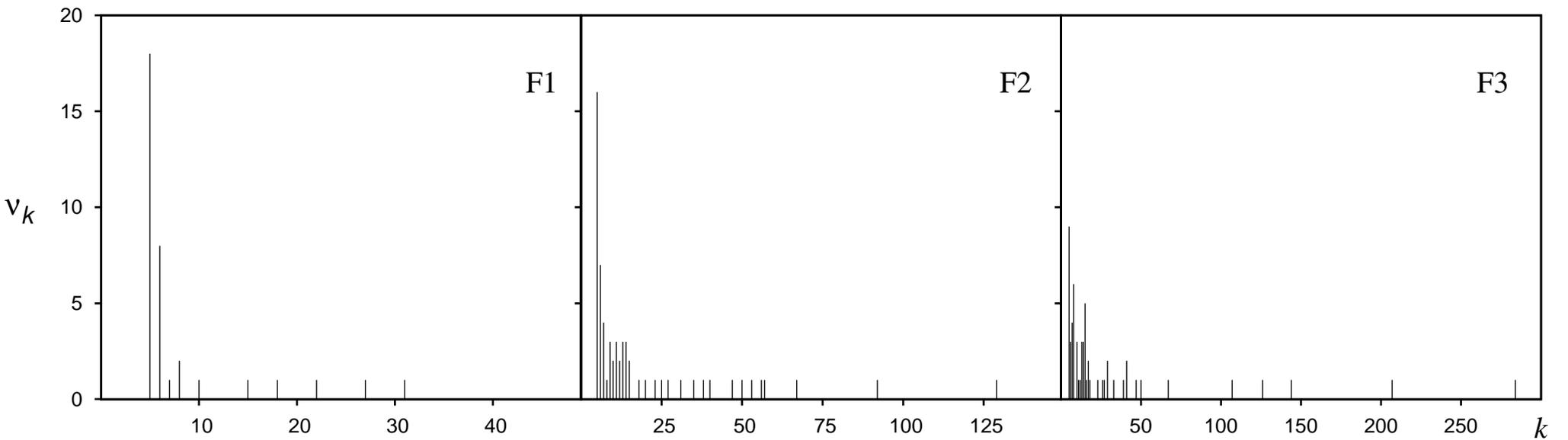
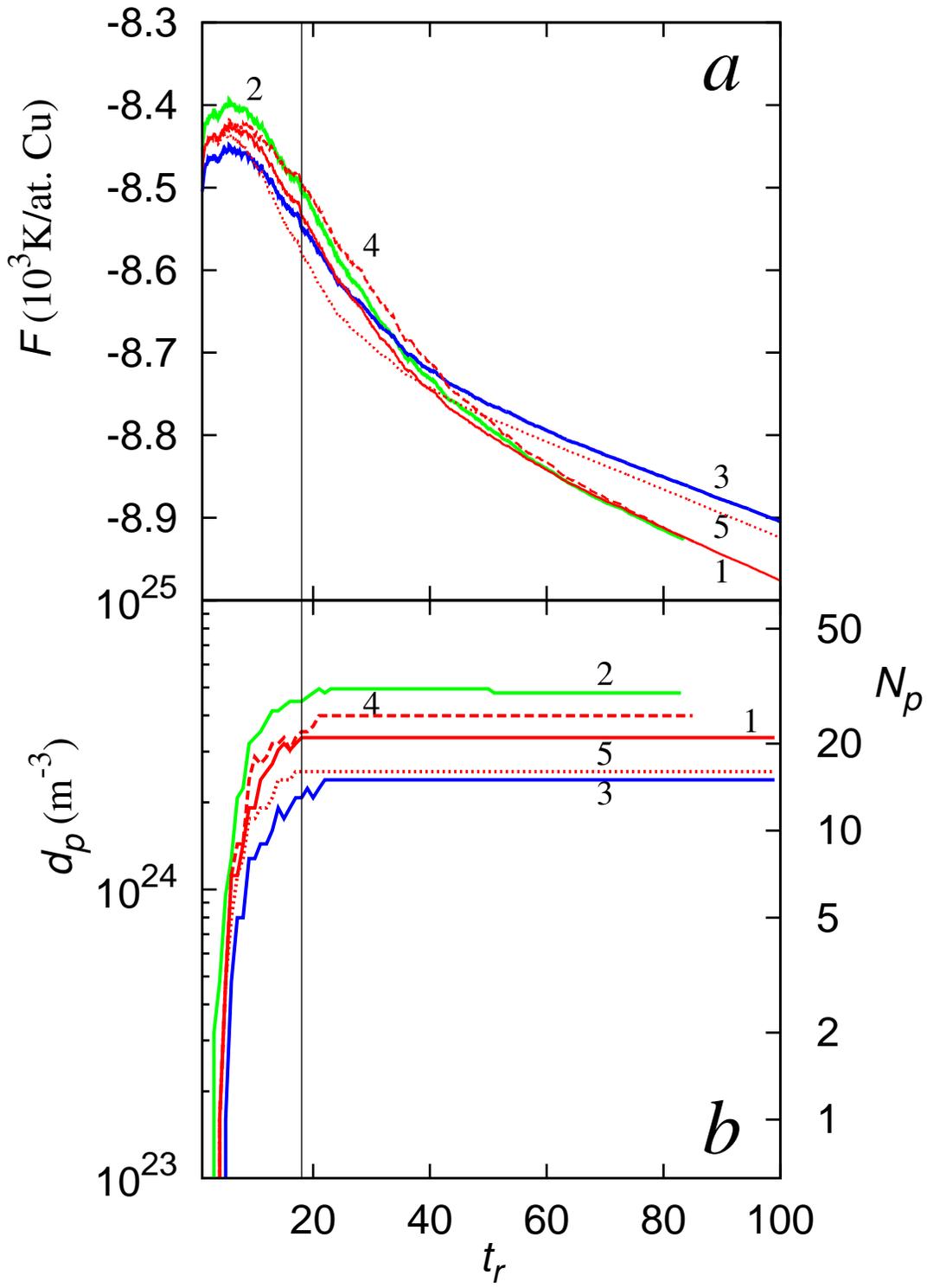
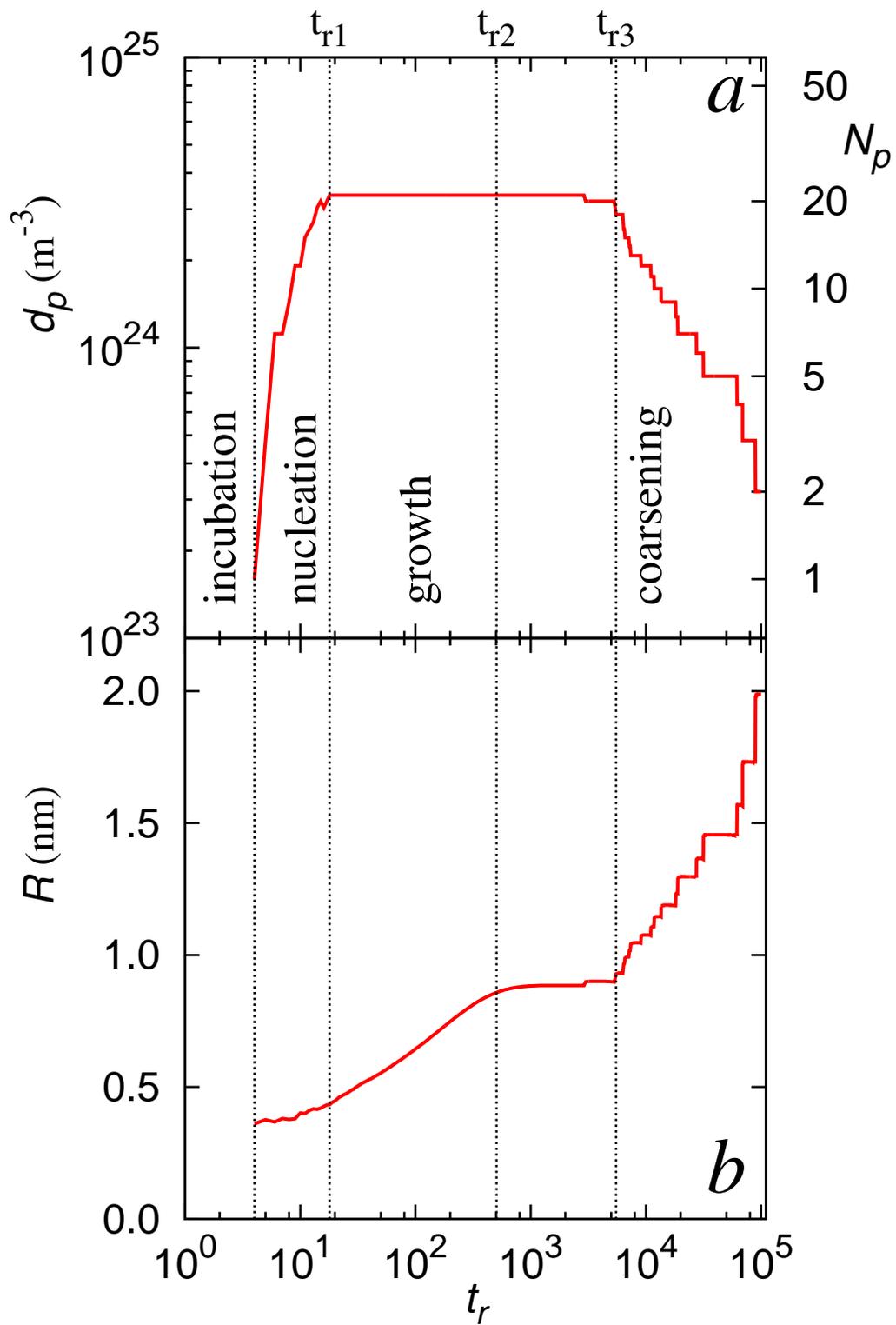
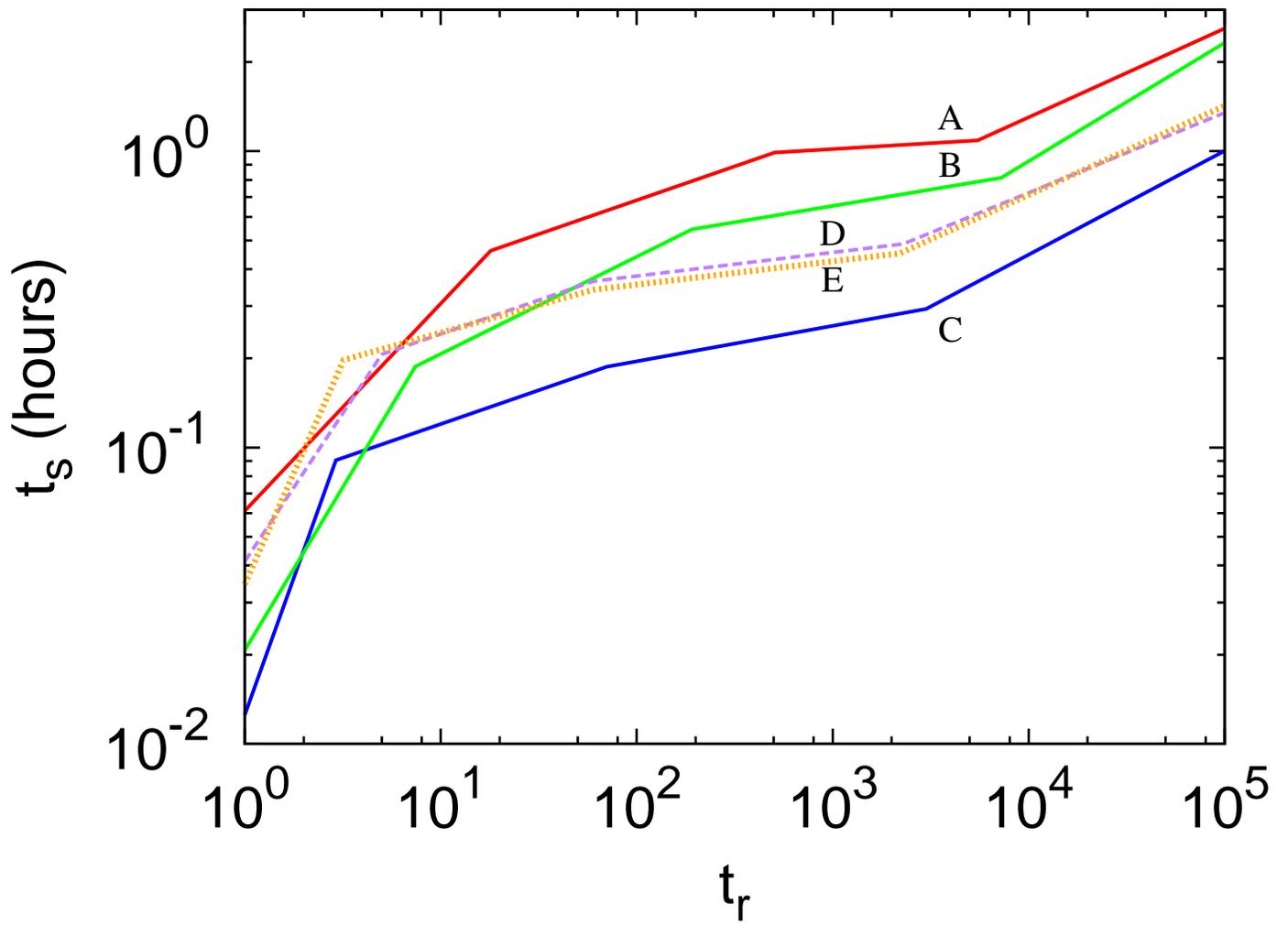
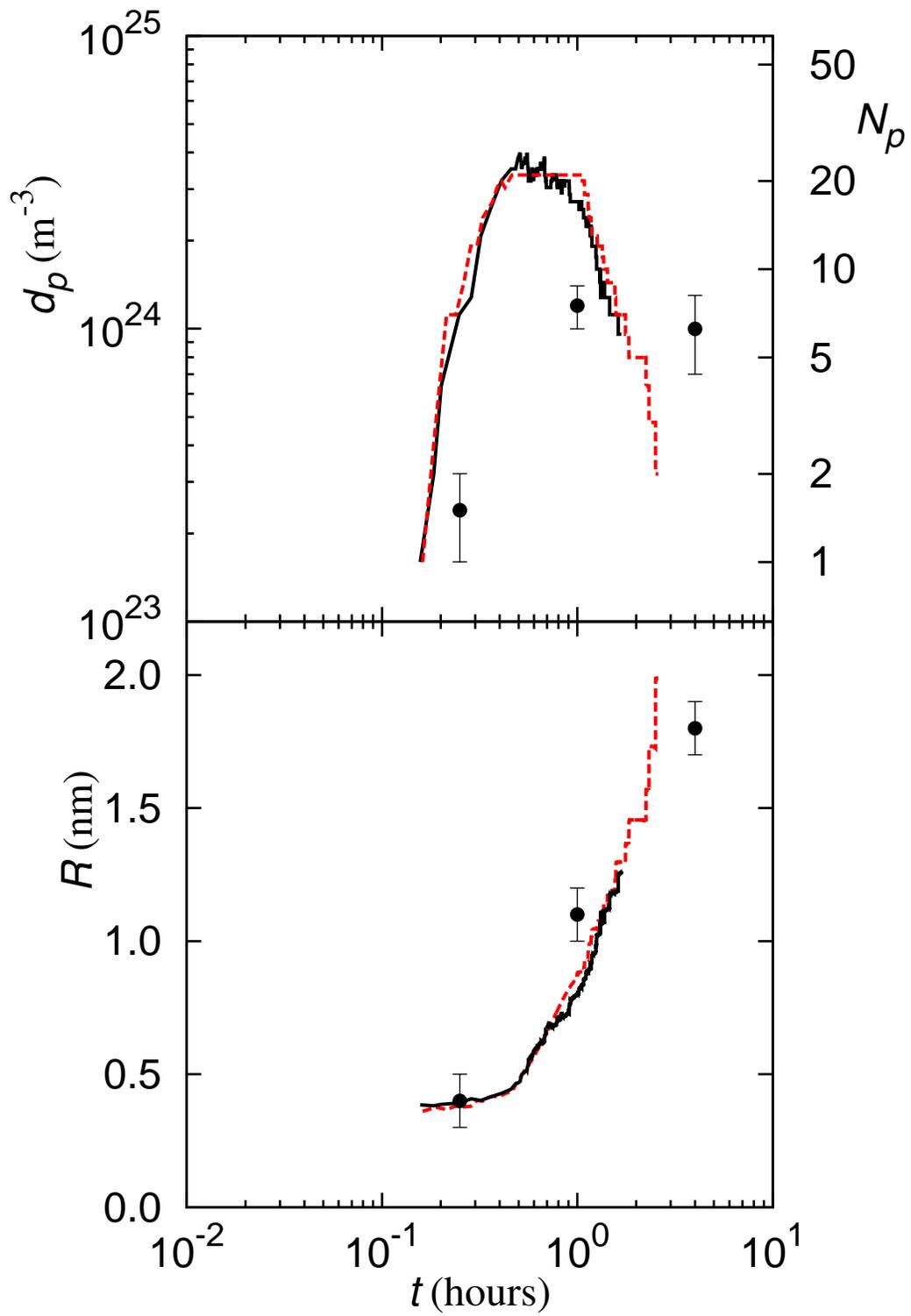
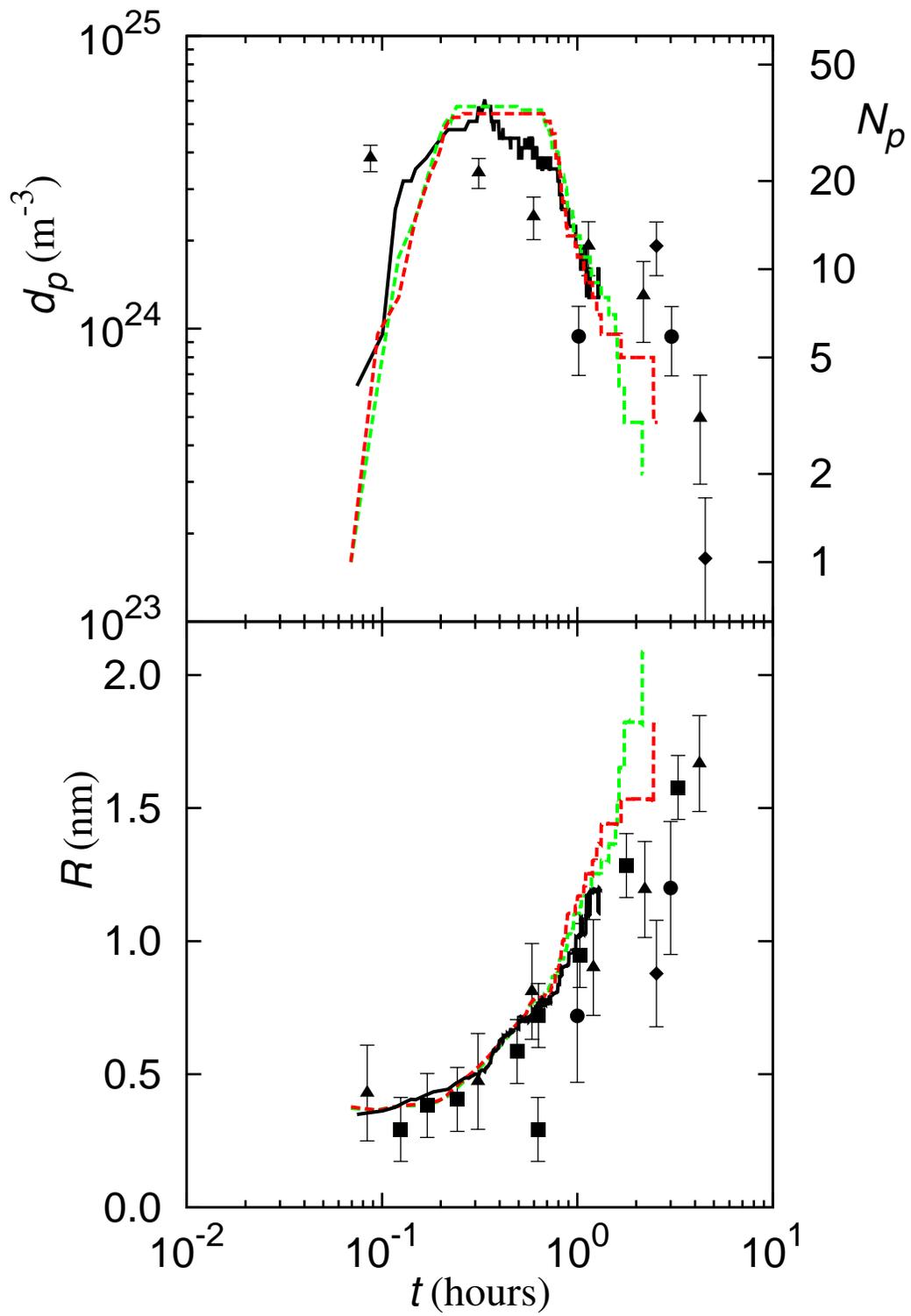
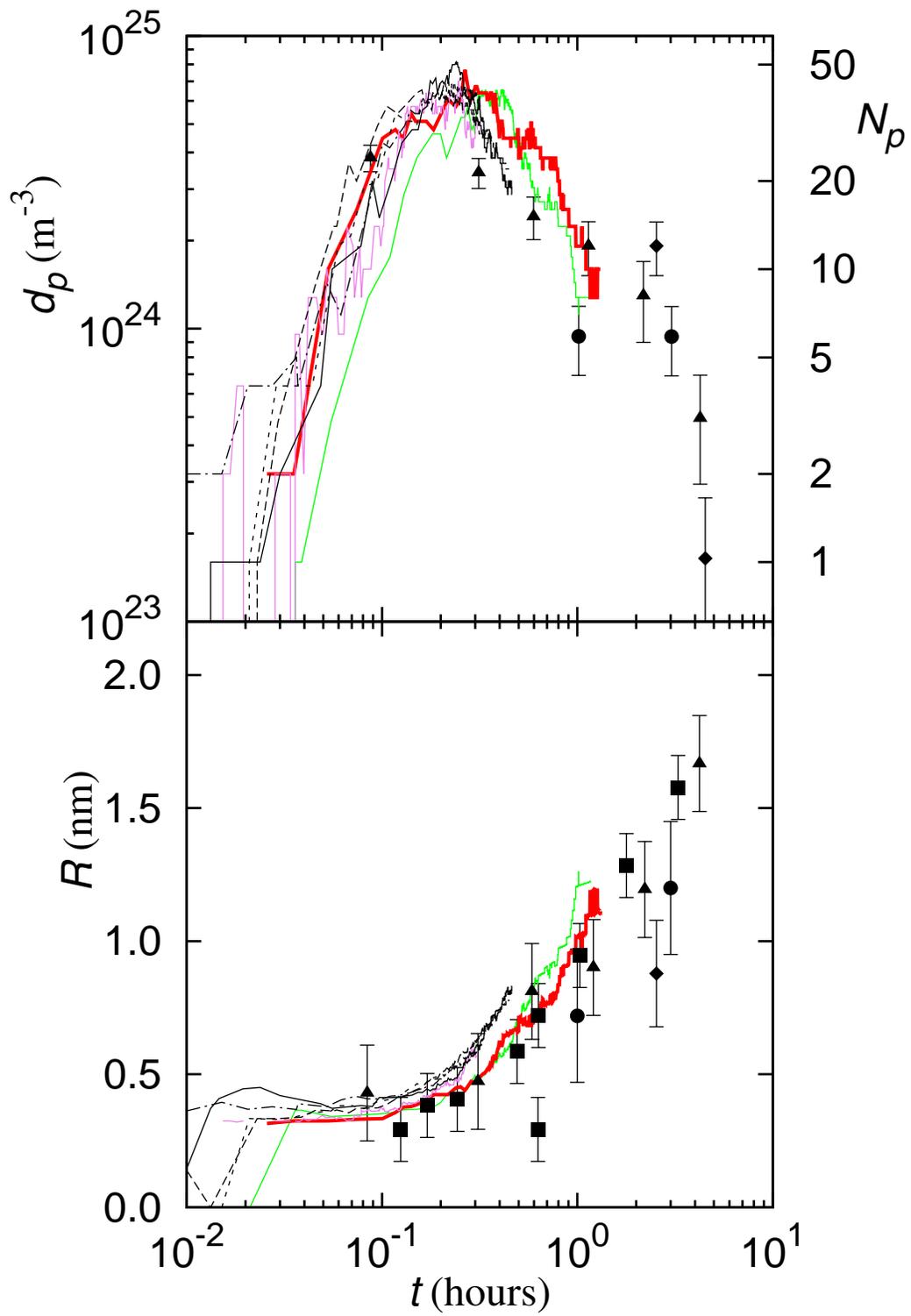
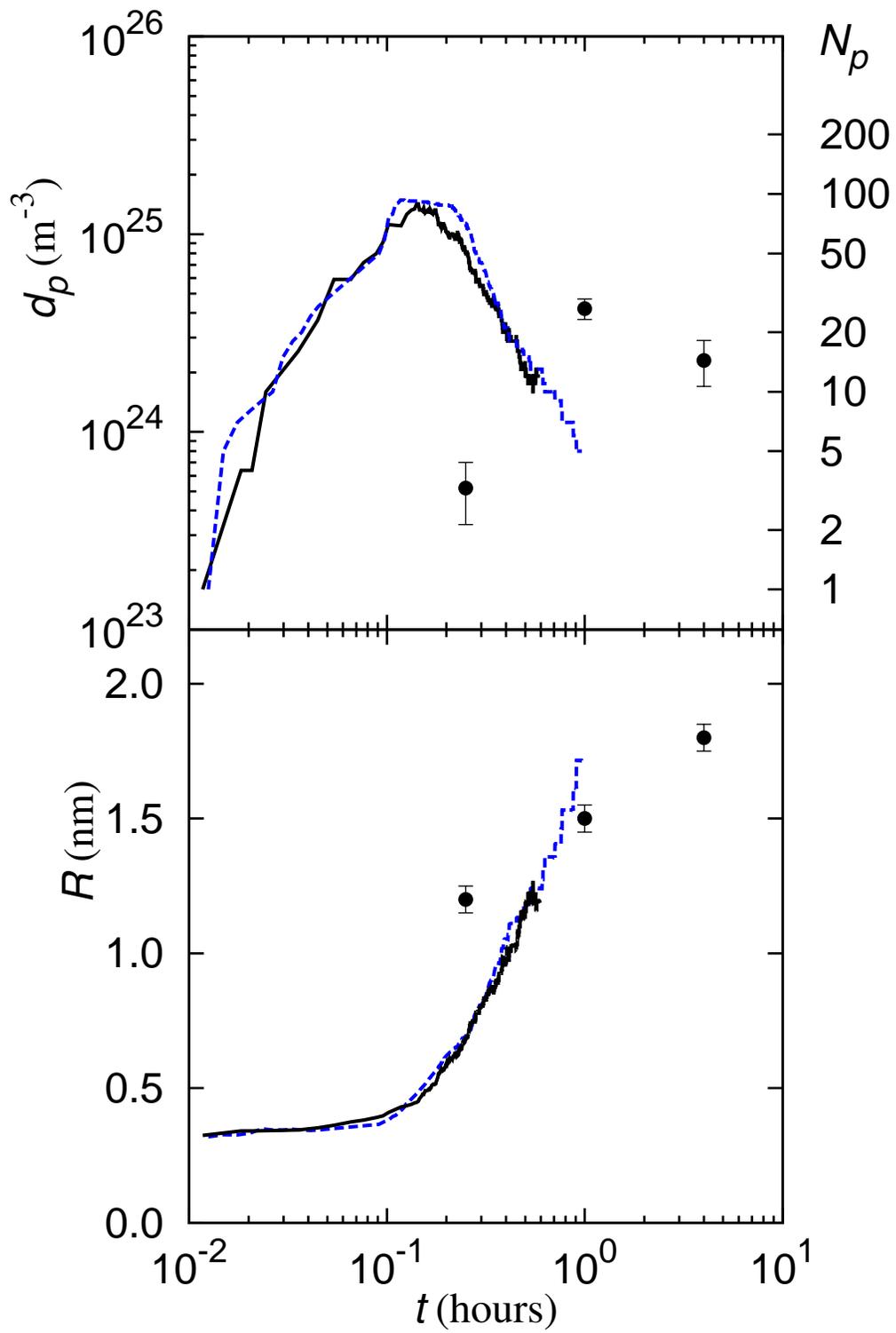
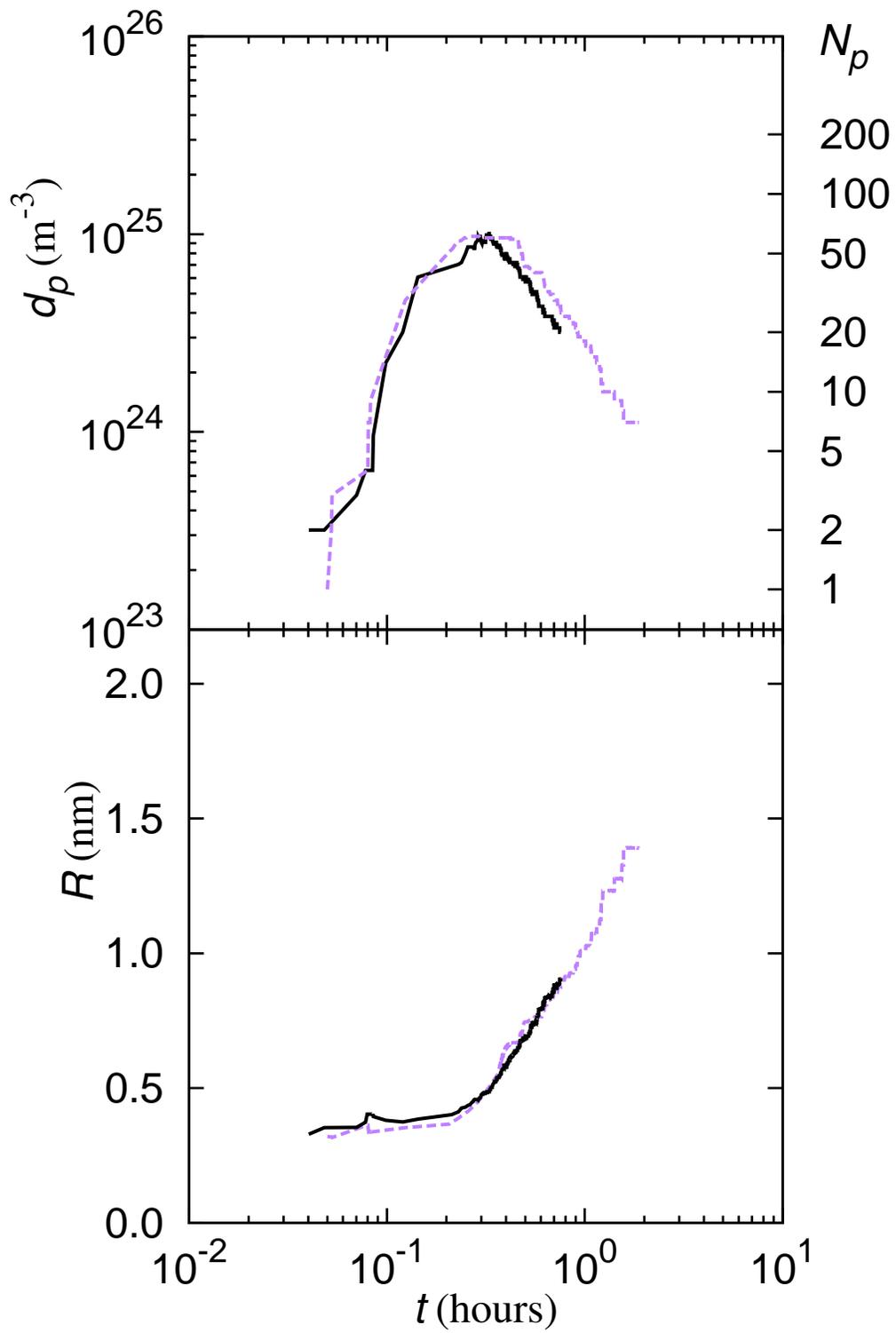
\end{document}